\documentclass{SciPost}

\usepackage[utf8]{inputenc}
\usepackage{amsfonts}
\usepackage{physics}
\usepackage{faktor}
\usepackage{graphicx}
\usepackage{tikz}
\usepackage{amsthm}
\numberwithin{equation}{section}
\usepackage[english]{babel}
\usepackage{mathtools}
\usepackage{nicefrac}
\usepackage{lscape}
\usepackage{textgreek}
\usepackage{enumerate}
\usepackage{dsfont}
\usepackage{cleveref}
\usepackage[english]{babel}
\usepackage{lineno}
\newcommand{\normord}[1]{\vcentcolon\mathrel{#1}\vcentcolon}

\providecommand{\vcentcolon}{\mathrel{\mathop{:}}}
\hypersetup{
    colorlinks,
    linkcolor={red!50!black},
    citecolor={blue!50!black},
    urlcolor={blue!80!black}
}

\usepackage[bitstream-charter]{mathdesign}
\DeclareSymbolFont{usualmathcal}{OMS}{cmsy}{m}{n}
\DeclareSymbolFontAlphabet{\mathcal}{usualmathcal}

\fancypagestyle{SPstyle}{
\fancyhf{}
\lhead{\colorbox{scipostblue}{\bf \color{white} ~SciPost Physics }}
\rhead{{\bf \color{scipostdeepblue} ~Submission }}

\fancyfoot[C]{\textbf{\thepage}}
}

\begin{document}

\pagestyle{SPstyle}

\begin{center}{\Large \textbf{\color{scipostdeepblue}{
Double-scaled bosonic and fermionic embedded ensembles, complex SYK, and the dual Hilbert space \\
}}}\end{center}

\begin{center}\textbf{
Jarod Tall and Steven Tomsovic 
}\end{center}

\begin{center}
  {Department of Physics and Astronomy, Washington State University,\\
Pullman, WA 99164-2814 USA}
\\[\baselineskip]
 \href{mailto:email1}{\small jarod.tall@wsu.edu}\,,\quad
 \href{mailto:email2}{\small tomsovic@wsu.edu}
\end{center}

\section*{\color{scipostdeepblue}{Abstract}}
\textbf{\boldmath{%
We derive the density of states and $2$- and $4$-point functions of embedded ensembles for both fermions and bosons in the double-scaled limit. It is shown the models are equivalent to the double-scaled Sachdev-Ye-Kitaev model, expanding the double-scaled universality class to include both fermionic and bosonic systems. The models can be solved by introducing the Wick product of non-commuting Gaussian random variables. We show that deriving the Wick product is sufficient for computing the density of states, and properties of the Wick product can be used to compute $n$-point functions directly in the energy basis. In this context, the Wick product is equivalent to normal ordering of $q$-oscillators, which leads to the duality between moments of double-scaled models and expectation values in the chord Hilbert space. By considering operator probes as a second set of oscillators, we extend the duality to compute $n$-point functions. Embedded ensembles are equivalent to complex SYK at fixed charge, and we show working directly with embedded ensembles streamlines the derivations.
}}

\vspace{\baselineskip}


\vspace{10pt}
\noindent\rule{\textwidth}{1pt}
\tableofcontents
\noindent\rule{\textwidth}{1pt}
\vspace{10pt}


\section{Introduction}
\label{sec:intro}
 Hamiltonians of quantum chaotic systems are often modeled by the Gaussian ensembles (GE) of random Hermitian matrices \cite{Mehta2004, Haake2010}. However, using the GE to describe a many-body quantum system describes an unphysical situation where all the particles interact simultaneously~\cite{French1970, BOHIGAS1971, Dyson1972}. Put differently, the random matrix describing the Hamiltonian is dense in the Fock basis with statistically independent matrix elements. In a physical many-body system there is likely to be a maximum body-rank of the Hamiltonian operators, often two-body, which leads to far fewer independent defining matrix elements and sparseness with many vanishing elements. The density of states, as a consequence of the assumption of statistical independence, follows the Wigner semicircle law, which is not observed in, for example, nuclear shell model calculations~\cite{French1970, BOHIGAS1971 }.  Embedded random matrix ensembles were introduced as a generalization of the GE to incorporate finite-body interactions characteristic of physical quantum systems~\cite{French1970, BOHIGAS1971, Wong1972}. The numerical calculations done in~\cite{French1970, BOHIGAS1971, Wong1972} found a transition of the density of states from Gaussian to semicircle as the body-rank of the Hamiltonian was increased from two-body to many-body, and a theoretical derivation was given in~\cite{Mon1975}. Embedded ensembles have also been shown numerically to exhibit level repulsion and have level density fluctuations indistinguishable from the GE~\cite{Wong1972}, making them a paradigmatic model of many-body quantum chaos.  
 
 The $p$-body Hamiltonian of the embedded Gaussian orthogonal ensemble (EGOE) was first studied analytically in the seminal paper by Mon and French~\cite{Mon1975}. The Hilbert space is defined by distributing $m$ particles over $N$ sites. For fermions this gives a dimensionality of $D_f=\binom{N}{m}$ and for bosons $D_b=\binom{N+m-1}{m}$. The ensemble is thus defined by three parameters $(p,m,N)$, and our interest is in the ``double-scaled'' limit,\footnote{This is really a ``triple-scaled'' limit but the term double-scaled is used to match the terminology used in the SYK literature.}
\begin{equation}
    N,m \to \infty, \hspace{.5cm} \frac{m}{N}= \text{fixed}, \hspace{.5cm} \frac{p^2}{m}= \text{fixed}.
\end{equation}
In the dilute limit, $m/N \to 0$, there is no difference between fermions and bosons, since the Pauli-exclusion principle is effectively turned off. It was known very early on how to solve the model in the limit $p^2/m \to 0$ in the bulk of the spectrum \cite{Mon1975, Brody1981, Draayer1977, French1987, Tomsovic1986}. We will show the transition from Gaussian to semicircle is governed by the transition parameter,
\begin{align}
      q_{\pm} \equiv  \exp\left\{-\frac{p^2}{m}\frac{1}{1\pm\frac{m}{N}}\right\}
\end{align}
where the $-$ sign is for fermions and the $+$ sign is for bosons. The focus of our paper is on single-trace quantities, i.e. secular quantities \cite{Wong1972}.  For secular quantities in the double-scaled limit, there is no difference between the EGOE and the unitary case (EGUE), and we will work with the latter. We do not compute multi-trace quantities, which are important for understanding level density fluctuations.  

Denoting a product of fermionic annihilation operators as $\psi_{p,I} = \psi_{i_1}\dots \psi_{i_p}$, the EGUE$(p)$ for fermions is
\begin{equation}
\label{fegue0}
H = \sum_{I,L}J_{I,L}\psi^{\dagger}_{p,I}\psi_{p,L}
\end{equation}
where $J_{I,L}$ is a random matrix drawn from the GUE to be further specified in Sec.~\ref{1}. A significant body of research exists involving applications of embedded ensembles and developing analytic methods to understand properties of  Eq.~\eqref{fegue0}. An excellent reference on embedded ensembles is the textbook by Kota~\cite{Kota2014}; for shorter reviews see~\cite{Benet2002, Kota2001, Kota:2018}, and the original papers by French and co-authors remain highly relevant~\cite{Mon1975, Brody1981, CHANG1971, Draayer1977, French1987 }.

The Sachdev-Ye-Kitaev (SYK) model \cite{Kitaev2016_1, Kitaev2016_2, Maldacena2016b} is a very similar model where the Hamiltonian is a product of $p$ Majorana fermions, typically defined with $p=4$\footnote{The similarities between SYK and embedded ensembles were also discussed in \cite{garcia_2016,Garcia_2017}.}. The model is quantum chaotic, as shown from fluctuation properties in agreement with random matrix theory \cite{garcia_2016, Cotler2017} and computations of the quantum Lyapunov exponent \cite{Maldacena2015, Kitaev2016_2, Kobrin2020,Berkooz:2018jqr, Baur_2023}. The SYK model has a low temperature effective theory governed by the Schwarzian action \cite{Maldacena2016b, Kitaev2016_1, Kitaev2016_2}, which is the same action that describes Jackiw–Teitelboim (JT) gravity~\cite{Maldacena2016, Jensen2016, Mertens2023}, making SYK a very interesting model of holography. 

The SYK Hamiltonian does not have a global $U(1)$ symmetry and the Hilbert space is $2^N$-dimensional. The double-scaled limit of this model, i.e. DS-SYK, is defined by keeping $p^2/N=$ fixed, with a transition parameter, $q_{\text{SYK}}=\exp\{-2p^2/N\}$\cite{Cotler2017,Berkooz:2018jqr}. All of the interesting holographic properties of SYK, such as the maximally chaotic Lyupovanov exponent and the Schwarzian action, can be found in DS-SYK by taking $q_{\text{SYK}} \to 1$ while shifting to the edge of the spectrum \cite{Berkooz:2018jqr}. These properties were actually first discovered in a similar double-scaled $p$-body spin glass model \cite{Erdos2014, Berkooz:2018qkz}. It was shown in~\cite{Erdos2014} that the density of states of this model is a $q$-normal distribution, which interpolates between Gaussian and semicircle as $q$ goes from $1$ to $0$. Based on the results of \cite{Erdos2014} the $q$-normal distribution was shown to apply to DS-SYK in \cite{Cotler2017, Berkooz:2018jqr, Garcia_2017}. The $q$-normal distribution was also found in the EGUE for fermions and bosons in~\cite{Vyas2018 }. We extend the results of~\cite{Vyas2018} by giving an analytic proof of the $q$-normal distribution for the EGUE in the double-scaled limit (DS-EGUE) that does not rely on computing low-order moments or numerics. Moreover, our proof does not rely on the arguments of~\cite{Erdos2014}, contrary to the derivations for DS-SYK in \cite{Berkooz:2018jqr, Cotler2017,Garcia_2017}. 

In \cite{Berkooz:2018qkz, Berkooz:2018jqr, Berkooz:2024} a technique based on chord diagrams was developed for the $p$-spin model and DS-SYK to compute moments and $n$-point functions. Importantly, these techniques apply to any system that has the same chord diagram description and can thus be considered as a universality class \cite{Berkooz:2024}. We prove DS-EGUE, for fermions and bosons, has the same chord diagram description and is equivalent to DS-SYK with $q_{\text{SYK}} \to q_{\pm}$. This broadens the double-scaled universality class to include both fermionic and bosonic systems. Therefore, we answer the question discussed in \cite{Baldwin:2019, Swingle_2024} of whether or not a bosonic system can have the same holographic properties as SYK.\footnote{A bosonic model based on a Parisi hypercube construction was considered in \cite{Jia:2024tii}}

However, though it is shown that DS-EGUE has the same chord description as DS-SYK, chord diagrams are not used to solve the model. Instead, we develop new techniques to derive the density of states and compute $n$-point functions. Our derivations rely on defining the Wick product of non-commuting Gaussian random variables and proving it is equivalent to $q$-Hermite polynomials. We show how properties of the Wick product can be used to compute the $2$- and $4$-point functions directly in the energy basis. The chord diagram method of~\cite{Berkooz:2018qkz, Berkooz:2018jqr, Berkooz:2024} demonstrated a duality to the Hilbert space of $q$-oscillators, referred to as the chord Hilbert space. This duality is very important from the perspective of holography since the chord space is interpreted as the Hilbert space of the bulk gravitational Hamiltonian~\cite{Berkooz:2018qkz, Lin:2022rbf}. We show this duality also follows from the equivalence between the $q$-Wick product and normal ordering of $q$-oscillators without needing to consider chord diagrams. Furthermore, by introducing a second set of oscillators, it is shown that the duality can be extended to compute $n$-point functions.

Sachdev proposed the two-body EGUE, defined in a $2^N$-dimensional Hilbert space instead of a fixed particle space, as a holographic dual to charged black holes in AdS$_2$ \cite{Sachdev:2015efa}. This was based on earlier work of the Sachdev-Ye (SY) model \cite{sachdev_1993, Georges2001, Parcollet_1998} where applications to holography were discussed in \cite{Sachdev_2010a, Sachdev_2010b}.\footnote{The SY model is similar to spin generalizations of two-body embedded ensembles, which have been studied extensively \cite{Kota2014,Vyas2012}.} The EGUE($p$), again in a $2^N$ dimensional space, was further studied in \cite{Gu2020, Davison_2017, sachdev2021, Berkooz:2020 }, and referred to as the complex SYK model.\footnote{The terminology "complex SYK" is a bit unfortunate, since French and co-authors introduced embedded ensembles 20 years prior to the SY model and 40 years prior to the SYK model.} By definition, embedded ensembles are thus equivalent to a fixed charge sector of complex SYK, and our results show that complex SYK at fixed charge is exactly equivalent to SYK in the double-scaled limit. In~\cite{Berkooz:2020}, chord techniques, were used to derive the moments of fermionic complex DS-SYK in the $2^N$ space, and then the results were projected onto a fixed charge sector. The derivation was rather lengthy and the final result was not exactly equivalent to DS-SYK. We compare our results to those in~\cite{Berkooz:2020}, explain the differences, and show it is simpler to work directly in a fixed particle/charge sector, i.e. with embedded ensembles. 

An important purpose of this paper is to bridge the gap between the SYK and embedded ensembles literature, since much of what is being computed is the same, albeit with different techniques and terminology.

\subsection{Main results and outline}
 The main results of the paper are the following: 
 \begin{itemize}
     \item We show DS-EGUE, for both fermions and bosons, has the same chord diagram description as DS-SYK. Thus, the models are exactly equivalent with $q_{\text{SYK}} \to q_{\pm}$, where 
     \begin{align}
     \nonumber
  q_{\pm} =  \exp\left\{-\frac{p^2}{m}\frac{1}{1\pm\frac{m}{N}}\right\},
\end{align}
with the $+$ sign being for bosons and the $-$ sign being for fermions. This answers the question of whether or not a bosonic model can have the same properties as SYK.
\item We give a new derivation of the density of states by defining the Wick product of non-computing Gaussian random variables, i.e.~the $q$-Wick product, and proving it is equivalent to $q$-Hermite polynomials.
\item We compute the exact $2$- and $4$-point functions directly in the energy basis, i.e.~the strength density, $\ev{O^{\dagger}\delta(H-E_1)O\delta(H-E_2)}$, and the analogous $4$-point density. 
\item We show normal ordering of the $q$-oscillator transfer matrix is equivalent to the $q$-Wick product, leading to the duality with the chord Hilbert space. A second set of $q$-oscillators can then be used to compute $n$-point functions in the dual Hilbert space.
\item We compare our results for the fermionic case to prior work on complex SYK and show working with embedded ensembles streamlines the derivations.
 \end{itemize}
 In Sec.~\ref{1}, we introduce the EGUE for fermions and bosons, show how to compute the moments in fixed particle spaces, and derive the transition parameter. The model is proven to have the same chord diagram description as DS-SYK. In Sec.~\ref{new derivation}, the Wick product of non-commuting Gaussian random variables is introduced. We prove it is equivalent to $q$-Hermite polynomials, which is sufficient for proving the density of states is a $q$-normal distribution. In Sec.~\ref{strength}, the $2$- and $4$-point strength densities are introduced and we develop a technique to solve them. In Sec.~\ref{chord}, it is shown that normal ordering of $q$-oscillators is equivalent to the $q$-Wick product, which demonstrates the duality to the chord Hilbert space. We then show how this property can be used to compute the $2$- and $4$-point functions. In Sec.~\ref{complex}, the results are compared to previously derived results for fermionic complex SYK at fixed charge and the differences discussed. In Sec.~\ref{conclusion}, the techniques and results are summarized and some future research directions noted. In Appendix~\ref{appendixa}, we collect some useful $q$-formulas and go through the technical details of deriving the $4$-point function. In Appendix~\ref{review}, we review the chord diagram and transfer matrix technique. In Appendix~\ref{proofT}, the details are given for proving that the normal ordered product of transfer matrices are the $q$-Hermite polynomials.

\section{Double-scaled fermionic and bosonic EGUE}
\label{1}

 In this section, we introduce the fermionic and bosonic embedded ensembles and demonstrate how to compute the moments in the double-scaled limit. We show the moments can be written as a sum over chord diagrams where each intersection has a weighting of $q_{\pm}$. Contrary to prior derivations for other double-scaled models, our proof does not rely on the arguments of~\cite{Erdos2014}. The derivation essentially follows from the moment method introduced in~\cite{Mon1975} and unitary group decompositions of $p$-body operators introduced in~\cite{CHANG1971,Draayer1977}. An immediate corollary of the result is that the moments of DS-EGUE, for both fermions and bosons, are exactly equivalent to DS-SYK with $q_{SYK} \to q_{\pm}$.

\subsection{Fermions}
Consider $m$ fermions distributed across $N$ sites. The Hilbert space has dimension $D_f = \binom{N}{m}$ and the relevant operators are $\psi_i :\hspace{.1cm} i \in (1,\dots, N)$ with the standard anti-commutation relations

\begin{equation}
    \{\psi_i,\psi^{\dagger}_j\}=\delta_{i,j}, \hspace{.5cm}  \{\psi_i,\psi_j\}=  \{\psi^{\dagger}_i,\psi^{\dagger}_j\}=0.
\end{equation}
It is useful to define a product of fermion operators as
\begin{equation}
    \psi^{\dagger}_{p,I} = \psi^{\dagger}_{i_p}\dots\psi^{\dagger}_{i_1}, \hspace{.5cm} \psi_{p,I} = \psi_{i_1}\dots\psi_{i_p}
\end{equation}
where $I$ denotes the index set $I=\{i_1,\dots, i_p\}$ and $1 \leq i_1 < \dots < i_p \leq N $.
The basis vectors of the Hilbert space can be written as a product of $m$ creation operators acting on the ground state
\begin{equation}
    \ket{\mu} = \psi^{\dagger}_{\mu_m}\dots\psi^{\dagger}_{\mu_1}\ket{0} =\psi^{\dagger}_{m,\mu}\ket{0}.
\end{equation}
The Hamiltonian of the fermionic EGUE$(p)$ is a $p$-body operator defined by embedding a random matrix in the $m-$particle space,
\begin{equation}
\label{fegue}
H = \sum_{I,L}J_{I,L}\psi^{\dagger}_{p,I}\psi_{p,L}
\end{equation}
where $J_{I,L}$ are complex, Gaussian random variables, i.e. distributed according to the GUE\footnote{$J_{I,L}$ could also be taken from the Gaussian orthogonal ensemble (GOE), which would define the EGOE$(p)$ but in the double-scaled limit all results computed in this paper will be equivalent.}, with a second moment of
\begin{equation}
\label{eguenorm}
    \expval{J_{I,L}J_{I',L'}}_J = \binom{m}{p}^{-1}\binom{N-m+p}{p}^{-1} \delta_{I,L'}\delta_{L,I'}.
\end{equation}
It will be shown this specific normalization sets the variance of the Hamiltonians to one. It then follows that when $p=m$ the embedded ensembles reduce to the GUE,
\begin{equation}
\label{GUE}
    \bra{\nu}H\ket{\mu}|_{p=m}=J_{\nu,\mu}.
\end{equation}
The moments of $H$ are defined by taking the $m-$particle expectation value\footnote{The $m$-particle expectation value is the trace in the $m$-particle space normalized to unity.} and the ensemble average over $J$,
\begin{equation}
\label{moments}
    \expval{H^k} \equiv \expval{H^k}_{m,J}=  \binom{N}{m}^{-1}\sum_{\mu}\bra{\mu} \expval{H^k}_J\ket{\mu}.
\end{equation}
It is important to note that by defining the moments as expectation values and setting the variance to unity means  Eq.~\eqref{moments} defines the \textit{reduced moments} of the Hamiltonian. The reduced moments are the quantities that should be compared across different models, and it is these moments that are equivalent between DS-SYK and DS-EGUE.

\begin{figure}
        \centering\includegraphics[width=0.5\linewidth]{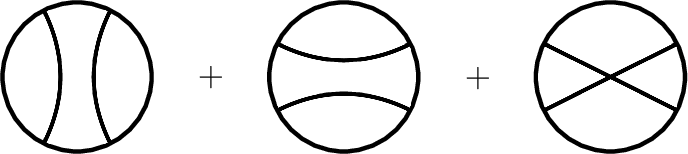}
        \caption{The three chord diagrams for the fourth order moment.}
        \label{fig1}
    \end{figure}

The following identities are necessary to compute the moments~\cite{Kota2014}: 
\begin{equation}
\begin{aligned}
\label{identities}
    & \sum_{I} \psi^{\dagger}_{p,I}\psi_{p,I}\ket{\mu} = \binom{m}{p}\ket{\mu} \hspace{.1cm}\rightarrow  \hspace{.1cm}\sum_{I} \psi^{\dagger}_{p,I}\psi_{p,I} = \binom{\hat{n}}{p}  \\[8pt]
   & \sum_{I} \psi_{p,I}\psi^{\dagger}_{p,I}\ket{\mu} = \binom{N-m}{p}\ket{\mu} \hspace{.1cm}\rightarrow  \hspace{.1cm}\sum_{I} \psi^{\dagger}_{p,I}\psi_{p,I} = \binom{N-\hat{n}}{p}
\end{aligned}
\end{equation}
where $\hat{n}$ denotes the number operator. The second moment can then be computed:
\begin{equation}
    \begin{aligned}
    \label{2nmoment}
        \expval{H^2} &=\binom{m}{p}^{-1}\binom{N-m+p}{p}^{-1} \sum_{I,J} \expval{\psi^{\dagger}_{p,I}\psi_{p,J}\psi^{\dagger}_{p,J}\psi_{p,I}}_m \\
        &= \binom{m}{p}^{-1}\binom{N-m+p}{p}^{-1} \sum_{I} \expval{\psi^{\dagger}_{p,I}\binom{N-\hat{n}}{p}\psi_{p,I}}_m \\
        &=\binom{m}{p}^{-1} \sum_{I} \expval{\psi^{\dagger}_{p,I}\psi_{p,I}}_m \\ 
        &=1.
    \end{aligned}
\end{equation}
We now turn to computing the fourth order moment which defines the transition parameter. The fourth order moment is actually known exactly for the fermionic embedded ensemble~\cite{Kota2014, Vyas2018}, however, the exact result is not necessary in the double-scaled limit. It is convenient to denote Wick contracted pairs by a common letter, e.g. $\expval{H^2}=\expval{AA}_m$. The fourth order moment is then a sum over three terms:
\begin{equation}
    \begin{aligned}
    \label{fourth}
        \expval{H^4} &= \expval{AABB}_m + \expval{ABBA}_m + \expval{ABAB}_m \\
        &= 2\expval{AABB}_m + \expval{ABAB}_m \\
        &= 2 + \expval{ABAB}_m
    \end{aligned}
\end{equation}
    where we used the cyclic invariance of the $m-$particle trace. This can be visually represented by a sum over three chord diagrams, see Fig.~\ref{fig1}. Thus, it remains to evaluate the intersection term $ABA$, which can be thought of as the intersection of two chords.  
    To do this, note that a generic $p$-body operator, e.g. the Hamiltonian, has a $U(N)$ symmetry. In general the operator does not transform irreducibly, but can be decomposed into irreducible representations of $U(N)$~\cite{CHANG1971}, which corresponds to the body rank of the operator. A reducible $p$-body operator can then be written as a sum over $s$-body irreducible representations with $s\le p$~\cite{CHANG1971, Draayer1977, Kota2014}. For example,
    \begin{equation}
        \begin{aligned}
        \label{reduce}
            B= \sum_{s=0}^{p}  B^{(s,p)}_{K,L}\psi^{\dagger}_{s,K}\psi_{s,L},
        \end{aligned}
    \end{equation}
    where the sum over $(K,L)$ has been made implicit. The importance of this decomposition is that the irreducible operators have particle-hole symmetry, $ B^{(s,p)}_{K,L} \psi^{\dagger}_{s,K}\psi_{s,L} = (-1)^sB^{(s,p)}_{K,L} \psi_{s,L}\psi^{\dagger}_{s,K}$. The operator $ABA$ can then be solved for, assuming it is always acting on an $m-$particle state:
   \begin{equation}
       \begin{aligned}
       \label{intersect}
           ABA &=\binom{m}{p}^{-1}\binom{N-m+p}{p}^{-1}  \sum_{I,J} \psi^{\dagger}_{p,I}\psi_{p,J}B\psi^{\dagger}_{p,J}\psi_{p,I} \\
           &=\binom{m}{p}^{-1}\binom{N-m+p}{p}^{-1}  \sum_{I,J,s}(-1)^s B^{(s,p)}_{K,L} \psi^{\dagger}_{p,I}\psi_{p,J} \psi_{s,L}\psi^{\dagger}_{s,K}\psi^{\dagger}_{p,J}\psi_{p,I} \\
           &=\binom{m}{p}^{-1}\binom{N-m+p}{p}^{-1}  \sum_{I,J,s}(-1)^s B^{(s,p)}_{K,L} \psi^{\dagger}_{p,I}\psi_{s,L}\psi_{p,J} \psi^{\dagger}_{p,J}\psi^{\dagger}_{s,K}\psi_{p,I} \\
           &=\binom{m}{p}^{-1}\binom{N-m+p}{p}^{-1}  \sum_{I,s}(-1)^s B^{(s,p)}_{K,L} \psi^{\dagger}_{p,I}\psi_{s,L}\binom{N-\hat{n}}{p}\psi^{\dagger}_{s,K}\psi_{p,I} \\
           &=\sum^{p}_{s=0}\binom{m}{p}^{-1}\binom{N-m+p}{p}^{-1}\binom{N-m+p-s}{p}\binom{m-s}{p} B^{(s,p)}_{K,L}\psi^{\dagger}_{s,K}\psi_{s,L}. 
       \end{aligned}
   \end{equation}
   To get the second line we used the particle-hole symmetry of irreducible operators, then made use of the identities,  Eq.~\eqref{identities}. In the double-scaled limit the dominant contribution is from $s \sim p$, so make the substitution $s\rightarrow p-s$, and evaluate the binomial coefficients:
   \begin{equation}
   \begin{aligned}
      & \binom{m}{p}^{-1}\binom{N-m+p}{p}^{-1}\binom{N-m+s}{p}\binom{m-p+s}{p} = \\
      &= \prod_{j=0}^{p-1} \frac{(N-m+p-(p-s)-j)(m-(p-s)-j)}{(N-m+p-j)(m-j)} \\
      &= \prod_{j=0}^{p-1}\left(1-\frac{p-s}{N-m+p-j}\right)\left(1-\frac{p-s}{m-j}\right)\\
      &\sim \left(1-\frac{p}{N-m}\right)^p\left(1-\frac{p}{m}\right)^p \sim \exp\left\{\frac{-p^2}{m}\frac{1}{1-\frac{m}{N}}\right\} = q_{-}.
   \end{aligned}
   \end{equation}
   Therefore, Eq.~\eqref{intersect} becomes in the double-scaled limit
   \begin{equation}
       ABA = q_{-}\sum_s B^{(s,p)}_{K,L}\psi^{\dagger}_{s,K}\psi_{s,L} = q_{-}B,
   \end{equation}
   and the fourth order moment Eq.~\eqref{fourth} in the double-scaled limit is
   \begin{equation}
       \expval{H^4} = 2+q_{-}.
   \end{equation}
   To evaluate higher order moments, it is necessary to compute higher order intersections, i.e. $AB_1 \dots B_k A$. The product $B_1\dots, B_k$ is a reducible $kp-$body operator, and can thus be written as a sum over irreducible operators in the same way as Eq.~\eqref{reduce},
   \begin{align}
       B_1,\dots, B_k = \sum_{s=0}^{pk}B^{(s,pk)}_{K,L}\psi^{\dagger}_{s,K}\psi_{s,L}
   \end{align}
   where again the sum over $(K,L)$ is implicit. Following the exact same steps as in Eq.~\eqref{intersect} gives
   \begin{equation}
       \begin{aligned}
       \label{intersect_k}
       AB_1\dots B_{k} A   &=\sum^{pk}_{s=0}\binom{m}{p}^{-1}\binom{N-m+p}{p}^{-1}\binom{N-m+p-s}{p}\binom{m-s}{p} B^{(s,pk)}_{K,L}\psi^{\dagger}_{s,K}\psi_{s,L},
       \end{aligned}
   \end{equation}
   and making the substitution $s\to kp-s$, the binomial coefficients become
 \begin{equation}
   \begin{aligned}
      & \binom{m}{p}^{-1}\binom{N-m+p}{p}^{-1}\binom{N-m+p-(kp-s)}{p}\binom{m-kp+s}{p} \\[8pt]
      &\sim (q_{-})^{k}.
       \end{aligned}
   \end{equation}
  An important assumption here is that $k$ is finite, since $s \sim k$ so for large $k$ this breaks down. This result then gives
  \begin{align}
  \label{intersections}
      AB_1\dots B_{k} A = \left(q_{-}\right)^k B_1\dots B_k.
  \end{align}
  All higher order moments can thus be calculated by counting intersections. For example, consider the following term in the sixth moment, which has three intersections:
  \begin{equation}
      \expval{H^6} \supset \expval{ABCABC}_m = (q_{-})^2\expval{BCBC}_m = (q_{-})^3.
  \end{equation}
  The solution for arbitrary moments can be written as a sum over chord diagrams where each chord diagram represents a possible set of Wick contractions. Each intersection of the chord diagrams gives a factor of $q_{-}$, so the solution is, for finite $k$,
  \begin{align}
  \label{chords}
      \expval{H^{2k}} = \sum_{\text{chord diagrams w/
     $k$ chords}} \left(q_{-}\right)^{\text{\# of intersections}}.
  \end{align}
This is the exact same result that was derived for quantum spin glasses and the SYK model in the double-scaled limit \cite{Erdos2014, Cotler2017,Garcia_2017, Berkooz:2018jqr, Berkooz:2018qkz}. The derivation for those models relied on probabilistic arguments about intersections of sets \cite{Erdos2014}, while the result here essentially follows from the moment method introduced in \cite{Mon1975} and unitary group decompositions of $p$-body operators introduced in \cite{CHANG1971, Draayer1977}.

The sum in  Eq.~\eqref{chords} can be evaluated and the density of states can be derived using the chord space and transfer matrix method developed in \cite{Berkooz:2018jqr, Berkooz:2018qkz ,Berkooz:2024}, which we review in Appendix \ref{review}, or by use of the Riordan-Touchard formula \cite{Garcia_2017, Erdos2014, riordan_1975, Touchard_1952}. However, in Sec.~\ref{new derivation} we will give a new derivation for the density of states that does not rely on the chord space or the Riordan-Touchard formula. Moreover, this technique will be necessary for computing $n$-point functions in the energy basis.
\subsection{Bosons}
Now consider $m$ bosons acting on $N$ sites. The Hilbert space for bosons has a dimensionality of $D_{b}=\binom{N+m-1}{m}$. It is important to note the particle density for bosons, $\frac{m}{N}$, can be infinite, i.e it ranges from $(0, \infty)$. This has the potential for new physics compared to the fermionic case. The bosonic operators acting on the $N$ sites are labeled as $\chi_i \hspace{.1cm} : i\in (1,\dots,N)$ and they satisfty the standard commutation relations,
\begin{equation}
    [\chi_i,\chi^{\dagger}_j]=\delta_{i,j}, \hspace{.5cm}  [\chi_i,\chi_j]=  [\chi^{\dagger}_i,\chi^{\dagger}_j]=0.
\end{equation}
products of bosonic operators are defined as
\begin{equation}
    \chi_{p,I}=\prod_{i=1}^{N} \frac{\chi_i^{p_i}}{\sqrt{p_i!}}, \hspace{.5cm} \sum_{i=1}^N p_i = p,
\end{equation}
where the occupation number of the $i^{th}$ site is defined as $p_i$, and $I$ denotes the set of occupation numbers. The basis vectors of the Hilbert space are
\begin{equation}
    \ket{\mu}=\prod_{i=1}^N \frac{(\chi^{\dagger}_i)^{m_i}}{\sqrt{m_i!}}\ket{0}, \hspace{.5cm} \sum_{i=1}^N m_i = m.
\end{equation}
The Hamiltonian can be defined as a constrained sum over occupation numbers,
\begin{equation}
  H = \sum_{\substack{p_1+\dots+p_N=p\\p'_1+\dots+p'_N=p}} J_{I,L} \prod_{i=1}^{N}\frac{(\chi_i^{\dagger})^{p_i}}{\sqrt{p_i!}} \frac{\chi_i^{p'_i}}{\sqrt{p'_i!}} = \sum_{I,L}J_{I,L}\chi^{\dagger}_{p,I}\chi_{p,L}
\end{equation}
where $J_{I,L}$ is a GUE random matrix with a dimensionality of $\binom{N+p-1}{p}$ and second moment
\begin{equation}
    \expval{J_{I,L}J_{I',L'}}_J =  \binom{m}{p}^{-1}\binom{N+m-1}{p}^{-1} \delta_{I,L'}\delta_{L,I'}.
\end{equation}
which is chosen such that the second moment of the Hamiltonian is one. The bosonic embedded ensemble also reduces to the GUE when $p=m$,
\begin{equation}
    \bra{\nu}H\ket{\mu}|_{p=m} =J_{\nu,\mu}.
\end{equation}
The following identities are necessary for computing the moments,
\begin{equation}
\begin{aligned}
    &\sum_{I}\chi^{\dagger}_{p,I}\chi_{p,I}\ket{\mu}= \sum_{p_1+\dots+p_N=p}\prod_{i=1}^N\binom{m_i}{p_i}\ket{\mu} = \binom{m}{p}\ket{\mu} \hspace{.1cm} \rightarrow \sum_{I}\chi^{\dagger}_{p,I}\chi_{p,I}=\binom{\hat{n}}{p}  \\
    &\sum_{I}\chi_{p,I}\chi^{\dagger}_{p,I}\ket{\mu}= \sum_{p_1+\dots+p_N=p}\prod_{i=1}^N\binom{m_i+p_i}{p_i}\ket{\mu} = \binom{N+m+p-1}{p}\ket{\mu} \\
    &\rightarrow \sum_{I}\chi_{p,I}\chi^{\dagger}_{p,I}=\binom{N+\hat{n}+p-1}{p},
\end{aligned}
\end{equation}
where the Vandermonde identity was used in both lines. The second moment can then be solved for:
\begin{equation}
    \begin{aligned}
        \expval{H^2}&= \binom{m}{p}^{-1}\binom{N+m-1}{p}^{-1} \sum_{I,J}\expval{\chi^{\dagger}_{I,p}\chi_{J,p}\chi^{\dagger}_{J,p}\chi_{I,p}}_m
    \\
    &= \binom{m}{p}^{-1}\binom{N+m-1}{p}^{-1} \sum_{I}\expval{\chi^{\dagger}_{I,p}\binom{N+\hat{n}+p-1}{p}\chi_{I,p}}_m \\
    & =\binom{m}{p}^{-1}\sum_{I}\expval{\chi^{\dagger}_{I,p}\chi_{I,p}}_m \\
    &=1.
    \end{aligned} 
\end{equation}
To solve for higher order moments, the weight of intersections, i.e. the term $ABA$, needs to be computed. This follows exactly the same steps as in the fermion case, Eq.~\eqref{intersect}:
\begin{equation}
    \begin{aligned}
        ABA=\sum^{p}_{s=0}\binom{m}{p}^{-1}\binom{N+m-1}{p}^{-1}\binom{N+m+s-1}{p}\binom{m-s}{p} B^{(s,p)}_{K,L}\chi^{\dagger}_{s,K}\chi_{s,L}.
    \end{aligned}
\end{equation}
Making the substitution $s\rightarrow p-s$ and evaluating the binomial coefficients gives,
\begin{equation}
\begin{aligned}
    &\binom{m}{p}^{-1}\binom{N+m-1}{p}^{-1}\binom{N+m+(p-s)-1}{p}\binom{m-(p-s)}{p}
   \\
   & \sim \exp\left\{-\frac{p^2}{m}\frac{1}{1+\frac{m}{N}}\right\} = q_{+},
\end{aligned}
\end{equation}
so that in the double-scaled limit,
\begin{align}
    ABA = q_{+}B.
\end{align}
The same logic leads to a direct evaluation of $AB_1\dots B_k A$, as was done in Eq.~\eqref{intersect_k} for fermions,
\begin{align}
    A B_1 \dots B_k A = (q_{+})^k B_1 \dots B_k.
\end{align}
Therefore, computing moments of the bosonic embedded ensembles reduces to counting intersections of chord diagrams,
  \begin{align}
      \expval{H^{2k}} = \sum_{\text{chord diagrams w/ $k$ chords}} \left(q_{+}\right)^{\text{\# of intersections}}.
  \end{align}
  Again this result is sufficient for demonstrating that the reduced moments of bosonic DS-EGUE are equivalent to the reduced moments of DS-SYK. To conclude, the moments of both fermionic and bosonic DS-EGUE are
   \begin{align}
      \expval{H^{2k}}_{\pm} = \sum_{\text{chord diagrams w/ $k$ chords}} \left(q_{\pm}\right)^{\text{\# of intersections}}, \hspace{.5cm} q_{\pm} =  \exp\left\{-\frac{p^2}{m}\frac{1}{1\pm\frac{m}{N}}\right\},
  \end{align}
  where the $-$ sign is for fermions and the $+$ sign is for bosons.
\section{New derivation of the density of states}
\label{new derivation}
In this section we give a new derivation for the density of states of a double-scaled $p$-body model. We use $q\in(0,1)$ to denote a generic transition parameter, which could be that of DS-SYK, DS-EGUE, or any other $p$-body model that is equivalent in the double-scaled limit. The essential technique stems from the relationship between the Wick product and orthogonal polynomials.
\subsection{Wick product and its $q$-generalization}
\label{wick}
The Wick product is a way of defining a product of random variables such that the expectation value vanishes \cite{avram1987}.\footnote{The Wick product for Gaussian random variables is equivalent to normal ordering of creation and annihilation operators, see Section \ref{chord}.} Here, we will define the Wick product in a way suitable for the purposes of computing moments of non-commuting Gaussian random variables. The Wick product of $H^n$ is a polynomial of order $n$ in $H$:
\begin{align}
\label{poly}
    \normord{H^{n}} \,\, = \sum_{j=0}^{n} a_j H^j
\end{align}
where $\normord{H^{n}}$ denotes the Wick product of $H^n$. We define the Wick product for Gaussian random variables by the set of equations,
\begin{align}
\label{def}
    \ev{H^m \normord{H^{n}}} = c_n \delta_{n,m},\hspace{.5cm} m\leq n,
\end{align}
which are $n+1$ equations that determine the $n+1$ coefficients in Eq.~\eqref{poly}. A recursive definition of the Wick product can be derived by considering $ H\normord{H^n}$ and noting that it is a polynomial of order $n+1$. We then assume it can be expanded in terms of Wick products,
\begin{align}
    \label{expand}
    H\normord{H^n} \,\, = \sum_{j=0}^{n+1}b_j\normord{H^{j}}.
\end{align}
Applying Eq.~\eqref{def} for $m<n-1$,
\begin{equation}
    \Big \langle H^{m}H\normord{H^n} \Big \rangle =0 \hspace{.5cm} m<n-1
\end{equation}
immediately shows $b_j =0$ for $j<n-1$. To derive $b_{n-1}$, consider $m=n-1$, 
\begin{align}
\label{3.5}
    \Big \langle H^{n-1}H\normord{H^{n}} \Big \rangle = c_n = b_{n-1}c_{n-1} \rightarrow b_{n-1}= \frac{c_n}{c_{n-1}} .
\end{align}
The case $m=n$ is not interesting since odd moments vanish. Setting $m=n+1$ gives
\begin{equation}
\label{3.6}
    \Big \langle H^{n+1}H\normord{H^{n}} \Big \rangle  = b_{n+1}c_{n+1} + b_{n-1}\Big \langle H^{n+1}\normord{H^{n-1}} \Big \rangle  .
\end{equation}
It is then necessary to evaluate $\ev{H^{n+2}\normord{H^n}}$, which can be done by choosing two $H$'s in the $H^{n+2}$ to be contracted. An important point is that the remaining $H^n$ will have no more internal contractions since these will all give a zero contribution by the definition of the Wick product. For example, if the right most $H$ is chosen, summing over all contractions will give,
\begin{equation}
\begin{aligned}
    &H^{n}AA + H^{n-1}AHA + \dots +AH^{n}A =H^n\sum_{j=0}^{n}q^j = [n+1]_qH^{n},
\end{aligned}
\end{equation}
where $[n]_q$ is defined as\footnote{See Appendix \ref{qappendix} for a collection of $q$-formulas.},
\begin{equation}
  [n]_q \equiv  \frac{1-q^{n}}{1-q}.
\end{equation}
Choosing the second $H$ to the right will give a factor of $[n]_q$ and so on, so that
\begin{equation}
\begin{aligned}
\label{3.9}
    \Big \langle H^{n+2}\normord{H^{n}} \Big \rangle &= \Big \langle H^{n}\normord{H^{n}} \Big \rangle \sum_{k=1}^{n+1} \,[k]_q = c_n \sum_{k=1}^{n+1} \,[k]_q 
    \end{aligned}
\end{equation}
Rearranging Eq.~\eqref{3.6} and using Eq.~\eqref{3.5} and Eq.~\eqref{3.9} results in 
\begin{equation}
    b_{n+1} = \frac{c_n}{c_{n+1}}\left(\sum_{k=1}^{n+1}\,[k]_q -\sum_{k=1}^{n}\,[k]_q\right) =  \frac{c_n}{c_{n+1}} [n+1]_q.
\end{equation}
The Wick product then has the recursive formula,
\begin{equation}
    H\normord{H^{n}} \,\, = [n+1]_q\frac{c_n}{c_{n+1}}\normord{H^{n+1}} + \frac{c_n}{c_{n-1}}\normord{H^{n-1}}.
\end{equation}
Setting the normalization to
\begin{align}
    c_n = \frac{(q)_n}{(1-q)^n}= [n]_q!,
\end{align}
where the $q$-Pochhammer is defined as
\begin{equation}
    (a)_n\equiv(a;q)_n\equiv\prod_{i=0}^{n-1}(1-aq^{i}),
\end{equation}
shows the Wick product is equivalent to $q$-Hermite polynomials, $\normord{H^n}\,\,=\mathbf{He}_n(H,q)$, which satisfy the recursion
\begin{align}
    \mathbf{He}_{n+1}(H|q) = (H)\mathbf{He}_{n}(H|q) - [n]_q\mathbf{He}_{n-1}(H|q).
\end{align}
When $q=1$ the $q$-Hermite polynomials become the normal ones, $\mathbf{He}_n(H|q=1) =\mathbf{He}_n(H) $, and when $q=0$ they are the Chebychev polynomials, $\mathbf{He}_n(H|q=0) =\mathbf{U}_n(H/2) $. 
\subsection{Density of states}
The density of states can now be derived using the properties of orthogonal polynomials.
The definition of the Wick product Eq.~\eqref{def} can be written
 \begin{align}
    \Big \langle H^m \, \mathbf{He}_n(H|q) \Big \rangle = \delta_{n,m} [n]_q!,\hspace{.5cm} m\leq n.
\end{align}
so that
\begin{align}
  \label{ortho}
      \Bigg \langle \mathbf{He}_{n}(H|q) \mathbf{He}_{m}(H|q) \Bigg \rangle =  \delta_{n,m}\Bigg\langle H^n \, \mathbf{He}_{n}(H|q) \Bigg \rangle =  [n]_q! \delta_{n,m}.
  \end{align}
  The result Eq.~\eqref{ortho} uniquely determines the density of states, since the expectation value can be expanded in a continuous energy basis,
  \begin{align}
      \Bigg \langle \mathbf{He}_{n}(H|q) \mathbf{He}_{m}(H|q) \Bigg \rangle = \int \, dE \ \rho(E|q) \,\, \mathbf{He}_{n}(E|q) \mathbf{He}_{m}(E|q) = [n]_q! \delta_{n,m},
  \end{align}
  and $\rho(E,q)$ can be read off from the orthogonality properties of $q$-Hermite polynomials \cite{ISMAIL1987},
  \begin{equation}
      \begin{aligned}
      \label{qnormal}
        \rho(E(\theta),q) = \abs{\left(e^{2i\theta}\right)_{\infty}}^2\left(q\right)_\infty\frac{\sqrt{1-q}}{4\pi\sin{\theta}},
      \end{aligned}
  \end{equation}
  where $E(\theta)=\frac{2\cos(\theta)}{\sqrt{1-q}}$ with  $E \in \left(-\frac{2}{\sqrt{1-q}},\frac{2}{\sqrt{1-q}}\right)$. The density of states when $q \to 1$ can be read off from the orthogonality properties of Hermite polynomials $\mathbf{He}_n(E)$,
  \begin{equation}
      \lim_{q \to 1} \rho(E,q) = \frac{1}{\sqrt{2\pi}}e^{-\frac{E^2}{2}}.
  \end{equation}
      and when $q=0$ from the properties of Chebyshev polynomials $\mathbf{U}_n(\frac{E}{2})$,
  \begin{equation}
  \lim_{q \to 0} \rho(E,q) = \frac{1}{2\pi}\sqrt{4-E^2}.
  \end{equation}
  The reduced moments of double-scaled $p$-body models are then,
    \begin{align}
    \expval{H^k} = \int_{-\frac{2}{\sqrt{1-q}}}^{\frac{2}{\sqrt{1-q}}} \, dE \ \rho(E,q) E^k. 
\end{align}

\section{Strength density and the $4$-point function}
\label{strength}
In this section we show how properties of the Wick product and orthogonal polynomials can be used to compute $n$-point functions of DS-EGUE directly in the energy basis. We explicitly compute the $2$- and $4$-point functions and show they agree with the DS-SYK results in \cite{Berkooz:2018jqr}.

The $2$-point function of the EGUE in the energy basis was originally studied in \cite{Draayer1977} by formulating a theory of strength distributions. The motivation was to understand transitions from an initial energy state $\ket{E_i}$, acted on by an operator $O$, to some final state $\ket{E_f}$. It was recognized in \cite{Draayer1977, French1987}, that to compute the transition strength or transition probability, i.e the $2$-point function, $\abs{\bra{E_f}O\ket{E_i}}^2$, it is necessary to evaluate $\ev{O^{\dagger}\delta(H-E_i)O\delta(H-E_f)}$, which is the strength density/distribution. This quantity can be directly evaluated by expanding the delta functions in terms of orthogonal polynomials and computing the moments \cite{French1987, Kota2014}. By computing low order moments, the strength density was shown to be a bivariate Gaussian distribution when $q\to1$ in~\cite{French1987, Tomsovic1986}, and more recently a bivariate $q$-normal distribution for generic $q$ in \cite{Vyas_2020, Kota2022}. We extend these results by developing a technique to compute the moments to all orders in the double-scaled limit for the $2$- and $4$-point density.

In the embedded ensemble literature, it is standard to take the operator $O$ to be a $t$-body embedded ensemble, EGUE(t), that is statistically independent from the Hamiltonian which is represented by EGUE(p). The reasoning for the statistical independence is that any part of the operator $O$ that is statistically correlated with $H$ can separated as $O = \alpha H + R$, where $R$ is statistically independent from $H$ \cite{Kota2014}. The off diagonal matrix element of $O$ in the energy basis are then completely determined by $R$. Therefore, as long as it is specified $E_i \neq E_f$ in $\abs{\bra{E_f}O\ket{E_i}}^2$ then $O$ and $R$ do not have to be distinguished. 

From a different perspective, it was argued in \cite{Berkooz:2018qkz} that taking $O$ to be a random operator independent from $H$ is the correct choice to understand black hole physics. The general argument being that the spectrum of a black hole near the horizon in nearly AdS$_2$ spacetimes can be described by $p$-body ensembles in the correct temperature and large $N$ regimes. The relevant operators to probe the spectrum should be made up of the same basic constituents of the theory, such as fermions or bosons, but general enough that it leads to universal results. This leads to considering the operators of the same statistical class as the Hamiltonian, whether that is DS-SYK, DS-EGUE, or something else.    
\subsection{Strength density}
\label{strength2}
We define $O$ to be represented by the EGUE(t). For fermions this is
\begin{equation}
O = \sum_{I,L}J^{O}_{I,L}\psi^{\dagger}_{t,I}\psi_{t,L}
\end{equation}
and $J^{O}_{I,L}$ are complex gaussian random variables with variance
\begin{equation}
    \expval{(J^{O \dagger})_{I',L'}\,J^{O}_{I,L}}_{J^{O}} =  \binom{m}{t}^{-1}\binom{N-m+t}{t}^{-1} \delta_{I,L'}\delta_{L,I'}.
\end{equation}
The bosonic case follows analogously. In the double-scaled limit the quantity $\frac{tp}{m}$ is held fixed, and we define the correlation coefficient in the double-scaled limit as
\begin{equation}
\label{cc}
    \rho \equiv \ev{O^\dagger H O H} = \exp\left\{-\frac{tp}{m}\frac{1}{1\pm \frac{m}{N}}\right\},
\end{equation}
where again the plus sign is for bosons and the minus sign is for fermions, and we are now averaging over both the $O$ and $H$ ensembles\footnote{This $\rho$ is not to be confused with the density of states $\rho(E|q)$}. The result,  Eq.~\eqref{cc}, follows from the exact same steps as was done for the fourth order moment in Sec.~\ref{1}. It is not necessary to consider $t$-body operators and more general operators consisting of $t_1$ creation operators and $t_2$ annihilation operators can be considered instead:
\begin{align}
    O = \sum_{I,L}J^{O}_{I,L}\psi^{\dagger}_{t_1,I}\psi_{t_2,L},
\end{align}
where
\begin{equation}
   \expval{(J^{O \dagger})_{I',L'}\,J^{O}_{I,L}}_{J^{O}} =  \binom{m}{t_2}^{-1}\binom{N-m+t_2}{t_1}^{-1} \delta_{I,L'}\delta_{L,I'}.
\end{equation}
Then the correlation coefficient in the double-scaled limit would be changed to 
\begin{align}
    \rho = \exp\left\{\frac{-t_2p}{m}\frac{1}{1 \pm \frac{m}{N}}\right\}\exp\left\{\frac{-p(t_2-t_1)}{(m\pm N)}\right\}.
\end{align}
In what follows we will not make specific reference to the form of $\rho$, but we will assume it is bounded from $(0,1)$.\footnote{For the bosonic case this means restricting $t_1$ and $t_2$ such that $t_2\geq t_1$. In the dilute limit this will not matter since the term $\exp\left\{\frac{-p(t_2-t_1)}{(m\pm N)}\right\}$ would go to one.} The strength density is defined as
\begin{equation}
\label{2}
    \ev{O^{\dagger}\delta(H-E_1)O\delta(H-E_2)}.
\end{equation} 
The Fourier transform of this object gives the time-dependent $2$-point function,
\begin{equation} 
    \ev{O^{\dagger}e^{-it_1 H}Oe^{-it_2H}}= \int \, d E_1 \, dE_2 \ e^{-it_1 E_1-it_2E_2} \ev{O^{\dagger}\delta(H-E_1)O\delta(H-E_2)}.
\end{equation}
The focus here is on introducing general techniques that allow  Eq.~\eqref{2} and the analogous $4$-point density to be evaluated directly. The delta functions can be expanded in terms of $q$-Hermite polynomials,
\begin{equation}
    \delta(H-E_1) = \sum_{n_1}\frac{1}{[n_1]_q!}\rho(E_1|q)\mathbf{He}_{n_1}(H|q)\mathbf{He}_{n_1}(E_1|q),
\end{equation}
so the strength density can be written
\begin{equation}
    \label{2_point1}
     \ev{O^{\dagger}\delta(H-E_1)O\delta(H-E_2)}= D
     \rho(E_1|q)\rho(E_2|q)O(E_1,E_2|\rho),
\end{equation}
where $D$ is the dimensionality of the Hilbert space\footnote{The dimensionality factor is here because $\rho(E|q)$ is the unit normalized density of states and $\ev{}$ is the normalized $m$-particle trace.} and we have defined the $2$-point function
\begin{equation}
\label{2point_2}
    O(E_1,E_2|\rho) = \sum_{n_1, n_2} \frac{\mathbf{He}_{n_1}(E_1|q)\mathbf{He}_{n_2}(E_2|q)}{[n_1]_q![n_2]_q!}\Big \langle O^{\dagger} \,\mathbf{He}_{n_1}(H|q) \,O \,\mathbf{He}_{n_2}(H|q) \Big \rangle.
\end{equation}
Before evaluating the $2$- and $4$-point function it is necessary to discuss properties of the Wick product that allow expectation values of the form appearing in  Eq.~\eqref{2point_2} to be computed. The essential property is that whenever $\mathbf{He}_{n_1}(H|q)$ appears in an expectation value it can never have any internal contractions that contribute, thus it can only contract with $H$'s occurring in other polynomials. The reason for this is that internal contractions of higher order terms in the polynomial will always cancel with lower order terms in the polynomial. This property follows from the definition of the Wick product which we repeat here,
\begin{align}
\label{def1}
    \Big \langle H^m \, \mathbf{He}_n(H|q) \Big \rangle = \delta_{n,m} [n]_q!,\hspace{.5cm} m\leq n.
\end{align}
As a simple example consider,
\begin{equation}
\label{example}
    \mathbf{He}_3(H|q) = H^3-(2+q)H.
\end{equation}
There are three ways that the $H^3$ term can contract with itself: $AAH + AHA+HAA = (2+q)H$, which cancels with the second term in  Eq.~\eqref{example}. This example follows from  Eq.~\eqref{def1} when $m=1,n=3$. Extending this logic, only the highest order term in $\mathbf{He}_n(H|q)$, i.e. $H^n$, can contribute to expectation values. Moreover all of the $H$'s in the remaining $H^n$ have to be uncorrelated, so it should be considered as a product of $n$ independent non-commuting Gaussian random variables. We will denote this product as $\{H^n\}$. In expectation values it is then valid to make the replacement $\mathbf{He}_n(H|q) \to \{H^n\} $. From the orthogonality of $q$-Hermite polynomials,
\begin{align}
     \Big \langle \mathbf{He}_{n}(H|q) \mathbf{He}_{m}(H|q) \Big \rangle =  \Big\langle \{H^n\} \, \{H^m\} \Big \rangle =  [n]_q! \delta_{n,m}.
\end{align}
It is also follows that in expectation values $O^{\dagger}\{H^n\}O = \rho^n\{H^n\}$.
 With these rules, the expectation value in the strength density  Eq.~\eqref{2point_2} is simply
 \begin{equation}
 \begin{aligned}
     \Big \langle O^{\dagger} \,\mathbf{He}_{n_1}(H|q) \,O \,\mathbf{He}_{n_2}(H|q) \Big \rangle =      \Big \langle O^{\dagger} \,\{H^{n_1}\} \,O \,\{H^{n_2}\} \Big \rangle = \rho^{n_1}\Big\langle \{H^{n_1}\} \, \{H^{n_2}\} \Big \rangle = \rho^{n_1}[n_1]_q!\delta_{n_1,n_2},\\
     \,
     \end{aligned}
 \end{equation}
 and  Eq.~\eqref{2point_2} becomes
 \begin{equation}
     \begin{aligned}
     \label{2point2}
         O(E_1,E_2|\rho) = \sum_{n_1} \frac{\rho^{n_1}\mathbf{He}_{n_1}(E_1|q)\mathbf{He}_{n_1}(E_2|q)}{[n_1]!}.
     \end{aligned}
 \end{equation}
 The sum can be evaluated \cite{szabowski2013},
 \begin{align}
 \label{2point_3}
     O(E_1(\theta_1),E_2(\theta_2)|\rho) = \frac{(\rho^2)_{\infty}}{(\rho e^{i(\theta_1-\theta_2)},\rho e^{-i(\theta_1-\theta_2)},\rho e^{i(\theta_1+\theta_2)},\rho e^{-i(\theta_1+\theta_2)})_{\infty}} ,
 \end{align}
 with $E_i(\theta_i)=\frac{2\cos(\theta_i)}{\sqrt{1-q}}$. The strength density,
 \begin{equation}
    \frac{1}{D} \ev{O^{\dagger}\delta(H-E_1)O\delta(H-E_2)}= 
     \rho(E_1|q)\rho(E_2|q)O(E_1,E_2|\rho)
 \end{equation}
 is a bivariate $q$-normal distribution. The result agrees with the $2-$point function computed with chord techniques in \cite{Berkooz:2018qkz, Berkooz:2018jqr}. For embedded ensembles, the bivariate $q$-normal distribution of the strength density was demonstrated by computing low order moments in \cite{Vyas_2020}, whereas here we have given an all orders proof valid in the double-scaled limit. 

 The form of the sum in  Eq.~\eqref{2point2} makes it very easy to take the limiting cases, since the $q$-Hermite polynomials are just replaced with either Hermite polynomials or Chebychev polynomials. When $q=1$ \cite{szabowski2013},
    \begin{equation}
   \lim_{q\to 1} \frac{1}{D} \ev{O^{\dagger}\delta(H-E_1)O\delta(H-E_2)}= 
    \frac{1}{2\pi\sqrt{1-\rho^2}} \exp\left\{-\frac{E_1^2+E_2^2-2\rho E_1 E_2}{1-\rho^2}\right\},
 \end{equation} 
 which is a bivariate normal distribution that was first derived for embedded ensembles in \cite{French1987, Tomsovic1986}. When $q=0$ \cite{szabowski2013},
  \begin{equation}
   \lim_{q\to 0}  \frac{1}{D} \ev{O^{\dagger}\delta(H-E_1)O\delta(H-E_2)}= 
    \frac{(1-\rho^2)\sqrt{4-E_1^2}\sqrt{4-E_2^2}}{4\pi^2\left((1-\rho^2)^2-\rho(1+\rho^2)E_1E_2+\rho^2(E_1^2+E_2^2)\right)}.
 \end{equation} 
 The strength density is of fundamental importance to understanding the off-diagonal matrix elements in the eigenstate thermalization hypothesis (ETH) in many-body quantum chaotic systems.
 For example, $q\to 1$ is the case relevant for physical quantum systems, which are typically governed by few-body interactions, and the strength density has exactly the form predicted by ETH. When $q \to 0$ the Hamiltonian is a Wigner-Dyson ensemble and is not physical. However, the strength density is a surprisingly non-trivial function of $E_1$ and $E_2$, which is contrary to the usual discussion of off-diagonal matrix elements when the eigenvectors of the Hamiltonian are Haar random unitaries \cite{D_Alessio_2016}. A more thorough discussion of embedded ensembles and ETH will be left for a follow up paper.
 \subsection{$4$-point function}
 In this section we show how to compute the $4$-point density,
\begin{equation}
\begin{aligned}
\label{4}
   &\frac{1}{D^3} \ev{O^{\dagger}\delta(H-E_1)O^{\dagger}\delta(H-E_2)O\delta(H-E_3)O\delta(H-E_4)} = \\
     &\rho(E_1|q)\rho(E_2|q)\rho(E_3|q)\rho(E_4|q)O(E_1,E_2,E_3,E_4|\rho),
\end{aligned}
\end{equation}
where the $4$-point function is
    \begin{equation}
    \begin{aligned}
        \label{4point1}
        &O(E_1,E_2,E_3,E_4|\rho)=\\[8pt]
        &\sum_{n_1, n_2,n_3,n_4} \frac{\mathbf{He}_{n_1}(E_1|q)\mathbf{He}_{n_2}(E_2|q)\mathbf{He}_{n_3}(E_3|q)\mathbf{He}_{n_4}(E_4|q)}{[n_1]_q![n_2]_q![n_4]_q![n_4]_q!} \Big \langle O^{\dagger} \,\{H^{n_1}\} \,O^{\dagger} \,\{H^{n_2}\}\,O \,\{H^{n_3}\} \,O \,\{H^{n_4}\}  \Big \rangle.
        \end{aligned}
    \end{equation}
    Before turning to the computation of the expectation value, we will need one more property of the product $\{H^n\}$. Say we would like to choose $k$ of the $H$'s in $\{H^{n}\}$ to contract with $k$ of the $H$'s in $\{H^{m}\}$. This can be done with the $q$-binomial coefficient:
\begin{align}
    \label{q-binomial}
    \{H^n\} \to \binom{n}{k}_q \{H^{n-k}\} \{A^k\}, 
\end{align}
 where the $k$ contracted $H$'s are denoted with $A^k$ and
 \begin{equation}
     \binom{n}{k}_q = \frac{[n]_q}{[n-k]_q[k]_q}.
 \end{equation}
 The expectation value has two terms:
\begin{equation}
    \begin{aligned}
        &\Big \langle O^{\dagger} \,\{H^{n_1}\} \,O^{\dagger} \,\{H^{n_2}\}\,O \,\{H^{n_3}\} \,O \,\{H^{n_4}\}  \Big \rangle = \\[8pt]
        &\Big \langle O_1^{\dagger} \,\{H^{n_1}\} \,O_2^{\dagger} \,\{H^{n_2}\}\,O_2 \,\{H^{n_3}\} \,O_1 \,\{H^{n_4}\}  \Big \rangle +\Big \langle O_1^{\dagger} \,\{H^{n_1}\} \,O_2^{\dagger} \,\{H^{n_2}\}\,O_1 \,\{H^{n_3}\} \,O_2 \,\{H^{n_4}\}  \Big \rangle.
    \end{aligned}
\end{equation}
The subscripts indicate which operators are being contracted. We will also denote $O_1^{\dagger}HO_1=\rho_1$ and  $O_2^{\dagger}HO_2=\rho_2$, but in the end $\rho_1=\rho_2=\rho$. The first term is referred to as the uncrossed $4$-point function and the second term is the crossed $4$-point function.
\subsubsection{Uncrossed $4$-point function}
The uncrossed $4$-point function can be solved for by first contracting the operators,
\begin{equation}
    \begin{aligned}
        \Big \langle O_1^{\dagger} \,\{H^{n_1}\} \,O_2^{\dagger} \,\{H^{n_2}\}\,O_2 \,\{H^{n_3}\} \,O_1 \,\{H^{n_4}\}  \Big \rangle= \rho_{1}^{n_4}\rho_2^{n_2}\Big \langle \{H^{n_1}\}  \,\{H^{n_2}\}\,\{H^{n_3}\} \,\{H^{n_4}\}  \Big \rangle 
    \end{aligned}.
\end{equation}
Then choose $k$ of the $H$'s in $\{H^{n_1}\}$ to contract with $k$ of the $H$'s in $\{H^{n_2}\}$ and $l$ of the $H$'s in $\{H^{n_3}\}$ to contract with $l$ of the $H$'s in $\{H^{n_4}\}$. Using Eq.~\eqref{q-binomial} while summing over $k$ and $l$ gives,
\begin{equation}
    \begin{aligned}
    \label{uncrossedO}
        \rho_{1}^{n_4}\rho_2^{n_2}\Big \langle \{H^{n_1}\}  \,\{H^{n_2}\}\,\{H^{n_3}\} \,\{H^{n_4}\}  \Big \rangle & = \rho_{1}^{n_4}\rho_2^{n_2}\sum_{k,l}\binom{n_1}{k}_q\binom{n_2}{k}_q\binom{n_3}{l}_q\binom{n_4}{l}_q \\
        & \times \Big \langle \{H^{n_1-k}\}\{A^k\}\{A^k\}\{H^{n_2-k}\}  \,\{H^{n_3-l}\}\{B^l\}\{B^l\}\{H^{n_4-l} \} \Big \rangle \\
        &= \rho_{1}^{n_4}\rho_2^{n_2}\sum_{k,l}\binom{n_1}{k}_q\binom{n_2}{k}_q\binom{n_3}{l}_q\binom{n_4}{l}_q [k]_q![l]_q! \\
        & \times \Big \langle \{H^{n_1+n_2-2k}\} \,\{H^{n_3+n_4-2l} \} \Big \rangle \\
        &= \rho_{1}^{n_4}\rho_2^{n_2}\sum_{k,l}\binom{n_1}{k}_q\binom{n_2}{k}_q\binom{n_3}{l}_q\binom{n_4}{l}_q [k]_q![l]_q! [n_1+n_2-2k]_q! \\
&\times\delta_{n_1+n_2-2k,n_3+n_4-2l}.
    \end{aligned}
\end{equation}
To get the second line we used $\{A^k\}\{A^k\} = [k]_q!$ and in the third line we used the fact that $\{H^{n_1-k}\}\{H^{n_2-k}\}$ have no more contractions with each other, and are thus completely uncorrelated, so they can be combined into $\{H^{n_1+n_2-2k}\}$. To evaluate the sum for the uncrossed $4$-point function,
 \begin{equation}
    \begin{aligned}
    &O_{uncrossed}(E_1,E_2,E_3,E_4|\rho_1,\rho_2)= \\[8pt]
    &\sum_{n_1, n_2,n_3,n_4} \frac{\mathbf{He}_{n_1}(E_1|q)\mathbf{He}_{n_2}(E_2|q)\mathbf{He}_{n_3}(E_3|q)\mathbf{He}_{n_4}(E_4|q)}{[n_1]_q![n_2]_q![n_4]_q![n_4]_q!} \Big \langle O_1^{\dagger} \,\{H^{n_1}\} \,O_2^{\dagger} \,\{H^{n_2}\}\,O_2 \,\{H^{n_3}\} \,O_1 \,\{H^{n_4}\}  \Big \rangle,
        \end{aligned}
    \end{equation}
make the transformation $n_{1,2} \to n_{1,2} +k$ and $n_{3,4} \to n_{3,4} + l$, which results in
\begin{equation}
\label{4point2}
    \sum_{n_1, n_2,n_3,n_4,k,l} \frac{\rho_{1}^{n_4+l}\rho_2^{n_2+k}\mathbf{He}_{n_1+k}(E_1|q)\mathbf{He}_{n_2+k}(E_2|q)\mathbf{He}_{n_3+l}(E_3|q)\mathbf{He}_{n_4+l}(E_4|q)[n_1+n_2]_q!}{[n_1]_q![n_2]_q![n_3]_q![n_4]_q! [k]_q![l]_q!}\delta_{n_1+n_2,n_3+n_4}.
\end{equation}
In Appendix \ref{uncrossed_appendix} we show that this is equal to 
\begin{align}
\label{uncrossed_final}
   O_{\text{uncrossed}}(E_1,E_2,E_3,E_4|\rho_1,\rho_2) = O(E_3,E_4|\rho_1)O(E_2,E_3|\rho_2)\frac{\delta(E_1-E_3)}{\rho(E_1|q)},
\end{align}
which is consistent with the DS-SYK result in \cite{Berkooz:2018jqr}. Note that $\rho_1=\rho_2=\rho$.
\subsubsection{Crossed $4$-point function}
The crossed $4$-point function requires solving the expectation value,
\begin{align}
    \Big \langle O_1^{\dagger} \,\{H^{n_1}\} \,O_2^{\dagger} \,\{H^{n_2}\}\,O_1 \,\{H^{n_3}\} \,O_2 \,\{H^{n_4}\}  \Big \rangle.
\end{align}
To contract the $O_1$ operators, the correlated terms between $\{H^{n_1}\} $ and $\{H^{n_2}\} $ have to be removed:
\begin{equation}
    \begin{aligned}
    \label{term1}
    O_1^{\dagger} \,\{H^{n_1}\} \,O_2^{\dagger} \,\{H^{n_2}\}\,O_1     &=  \sum_{k} \binom{n_1}{k}_q\binom{n_2}{k}_qO_1^{\dagger} \,\{H^{n_1-k}\} \{A^{k}\} \,O_2^{\dagger} \,\{A^k\}\{H^{n_2-k}\}\,O_1 \\
    & =\sum_{k} \binom{n_1}{k}_q\binom{n_2}{k}_q [k]_q! \,\rho_2^k  O_1^{\dagger} \,\{H^{n_1-k}\} \,O_2^{\dagger} \,\{H^{n_2-k}\}\,O_1 \\
    & =\rho_{12} \sum_{k} \binom{n_1}{k}_q\binom{n_2}{k}_q [k]_q! \,  \rho_2^k \rho_1^{n_1+n_2-2k} \,\{H^{n_1-k}\} \,O_2^{\dagger} \,\{H^{n_2-k}\}
    \end{aligned}
\end{equation}
where $\rho_{12}=\exp{t^2/m(1\pm m/N)}$ if $O$ is a $t$-body operator. In the first line we summed over all contractions between $\{H^{n_1}\} $ and $\{H^{n_2}\}$ so that $\{H^{n_1-k}\}$ and $\{H^{n_2-k}\}$ are uncorrelated, which allows the $O_1$ operators to be contracted. The remaining expectation value is  
\begin{align}
      \Big \langle  \,\{H^{n_1-k}\} \,O_2^{\dagger} \,\{H^{n_2-k}\}\,\{H^{n_3}\} \,O_2 \,\{H^{n_4}\}  \Big \rangle.
\end{align}
We then remove correlated terms between $\{H^{n_2-k}\}$ and $\{H^{n_3}\}$: 
\begin{align}
\label{term2}
    O_2^{\dagger} \,\{H^{n_2-k}\}\,\{H^{n_3}\} \,O_2 = \sum_l\binom{n_2-k}{l}_q\binom{n_3}{l}_q[l]_q!\rho_2^{n_2+n_3-k-2l}\{H^{n_2-k-l}\}\,\{H^{n_3-l}\},
\end{align}
which leaves 
\begin{align}
      \Big \langle  \,\{H^{n_1-k}\} \,\{H^{n_2-k-l}\}\,\{H^{n_3-l}\} \,\{H^{n_4}\}  \Big \rangle.
\end{align}
The product $\{H^{n_1-k}\}$ is still correlated with $\{H^{n_3-l}\}$, removing the correlated terms gives
\begin{align}
\label{term3}
    \{H^{n_1-k}\} \,\{H^{n_2-k-l}\}\,\{H^{n_3-l}\} = \sum_{j}\binom{n_1-k}{j}_q\binom{n_3-l}{j}_q[j]_q!\, q^{j(n_2-k-l)}   \{H^{n_1+n_2+n_3-2k-2l-2j}\}.
\end{align}
To get this result note that the product $   \{H^{n_1-k-j}\} \,\{H^{n_2-k-l}\}\,\{H^{n_3-l-j}\} $ is now completely uncorrelated and equals $\{H^{n_1+n_2+n_3-2k-2l-2j}\}$. The final term is,
\begin{align}
\label{term4}
      \Big \langle  \,\{H^{n_1+n_2+n_3-2k-2l-2j}\} \,\{H^{n_4}\}  \Big \rangle = \delta_{n_1+n_2+n_3-2k-2l-2j, n_4}[n_4]_q!.
\end{align}
Collecting all the factors from Eq.~\eqref{term1}, Eq.~\eqref{term2}, Eq.~\eqref{term3}, and Eq.~\eqref{term4} results in
\begin{equation}
    \begin{aligned}
    \label{crossedO}
          &\Big \langle O_1^{\dagger} \,\{H^{n_1}\} \,O_2^{\dagger} \,\{H^{n_2}\}\,O_1 \,\{H^{n_3}\} \,O_2 \,\{H^{n_4}\}  \Big \rangle =\\[8pt]
          &\rho_{12}\sum_{k,l,j}\binom{n_1}{k}_q\binom{n_2}{k}_q\binom{n_2-k}{l}_q\binom{n_3}{l}_q\binom{n_1-k}{j}_q\binom{n_3-l}{j}_q\,[k]_q![l]_q![j]_q![n_4]_q!\, \\
          &\times \rho_1^{n_1+n_2-2k}\rho_2^{n_2+n_3-2l}q^{j(n_2-k-l)}\delta_{n_1+n_2+n_3-2k-2l-2j, n_4}.
    \end{aligned}
\end{equation} 
To evaluate the sum, 
 \begin{equation}
    \begin{aligned}
        O_{crossed}(E_1,E_2,E_3,E_4|\rho_1,\rho_2)&=\sum_{n_1, n_2,n_3,n_4} \frac{\mathbf{He}_{n_1}(E_1|q)\mathbf{He}_{n_2}(E_2|q)\mathbf{He}_{n_3}(E_3|q)\mathbf{He}_{n_4}(E_4|q)}{[n_1]_q![n_2]_q![n_4]_q![n_4]_q!}\\
        &\times \Big \langle O_1^{\dagger} \,\{H^{n_1}\} \,O_2^{\dagger} \,\{H^{n_2}\}\,O_1 \,\{H^{n_3}\} \,O_2 \,\{H^{n_4}\}  \Big \rangle,
        \end{aligned}
    \end{equation}
    make the transformation $n_1 \to n_1+k+j$, $n_2\to n_2+k+l$, $n_3 \to n_3+j+l$, which leads to 
     \begin{equation}
    \begin{aligned}
        &O_{crossed}(E_1,E_2,E_3,E_4|\rho_1,\rho_2)=\\[8pt]
        &\rho_{12}\sum_{n_1, n_2,n_3,j,k,l} \frac{\mathbf{He}_{n_1+k+j}(E_1|q)\mathbf{He}_{n_2+k+l}(E_2|q)\mathbf{He}_{n_3+j+l}(E_3|q)\mathbf{He}_{n_1+n_2+n_3}(E_4|q)}{[n_1]_q![n_2]_q![n_3]_q![j]_q![k]_q![l]_q!} \\
        &\times \rho_1^{n_1+n_2+l+j}\rho_2^{n_2+n_3+k+j}q^{jn_2}.
        \end{aligned}
    \end{equation}
In Appendix \ref{crossed_appendix} we evaluate the sum and show,
\begin{equation}
    \begin{aligned}
    \label{crossedOfinal}
        &O_{crossed}(E_1,E_2,E_3,E_4|\rho_1,\rho_2) = \rho_{12} \times \\[8pt]
         &O(E_1,E_2|\rho_2)O(E_3,E_4|\rho_2)O(E_2,E_3|\rho_1) O(E_1,E_4|\rho_1)\frac{(\rho_1e^{-i(\theta_2+\theta_3)},\rho_2 \rho_1 e^{i(\theta_3\pm \theta_1)},\rho_2 \rho_1 e^{i(\theta_2\pm \theta_4)})_{\infty}}{(\rho_1 \rho_2^2 e^{i(\theta_2+\theta_3)},\rho_1^2)_{\infty}} \\
   &{}_8W_7\left(q^{-1}\rho_1\rho_2^2 e^{i(\theta_2+\theta_3)};\rho_1e^{i(\theta_3+\theta_2)},\rho_2e^{i(\theta_2 \pm \theta_1)},\rho_2e^{i(\theta_3 \pm \theta_4)};q,\rho_1 e^{-i(\theta_3+\theta_2)}\right),
    \end{aligned}
    \end{equation}
    where ${}_8W_7$ is defined in  Eq.~\eqref{Wdef} and here the $\pm$ notation means taking both signs, e.g. $(\rho_2 \rho_1 e^{i(\theta_3\pm \theta_1)})=(\rho_2 \rho_1 e^{i(\theta_3+ \theta_1)},\rho_2 \rho_1 e^{i(\theta_3- \theta_1)}). $\footnote{Not to be confused with $\pm$ corresponding to fermions or bosons.} The result is consistent with the DS-SYK result computed with chord techniques in \cite{Berkooz:2018jqr}.
 \section{The dual Hilbert space}
 \label{chord}
 The chord diagram approach to computing moments of double-scaled models developed in \cite{Berkooz:2018qkz,Berkooz:2018jqr,Berkooz:2024}, and reviewed in Appendix \ref{review}, demonstrated a duality to ground state expectation values of the transfer matrix in the Hilbert space of $q$-oscillators, referred to as the chord Hilbert space. The duality is very important from the perspective of holography, since the transfer matrix is interpreted as the Hamiltonian of the bulk gravitational theory \cite{Berkooz:2018qkz, Lin:2022rbf}. In this section, we show that normal ordering of the transfer matrix gives the $q$-Hermite polynomials\footnote{This is well known for the $q= 1$ case \cite{wurm2002}.}, equivalent to the $q$-Wick product of DS-Hamiltonians, which leads to the duality of the moments. Furthermore, by considering a second set of oscillators, we show that the duality can be extended to compute $n$-point functions. 
 \subsection{Moments and the transfer matrix}
The $q$-oscillators satisfy the $q$-commutator relation,
  \begin{align}
        [a,a^{\dagger}]_q=a a^{\dagger} - qa^{\dagger}a=1.
    \end{align}
 The dual Hilbert space is
 \begin{align}
     \mathcal{H}_a = \Big \{ \,\ket{n}\,  | \, n \in \mathbb{Z_+}\,\Big\}
 \end{align}
and the operators act on the Hilbert space as
\begin{align}
    a\ket{n}=[n]_q \ket{n-1}, \hspace{.5cm} a^{\dagger}\ket{n}=\ket{n+1},
\end{align}with the transfer matrix being,
\begin{align}
    T = a+ a^{\dagger}.
\end{align}
Normal ordering of $T^n$ is defined by ignoring the commutator and moving all of the $a$'s to the right so that the ground state expectation values vanishes:
\begin{align}
    \bra{0}\normord{T^n}\ket{0}=0.
\end{align}
For $m \leq n$,
\begin{align}
\label{normal_def}
    \bra{0}T^m\normord{T^n}\ket{0}= \bra{0}a^m (a^{\dagger})^n\ket{0}=[n]_q! \delta_{n,m}, \hspace{.2cm} m\leq n,
\end{align}
which follows since by the definition of $\normord{T^n}$ only the term $(a^{\dagger})^{n}$ in this polynomial could contribute.
The definition of $\normord{T^n}$ in Eq.~\eqref{normal_def} is equivalent to the $q$-Wick product definition of $\normord{H^n}$ in Eq.~\eqref{def}. Following the steps in \ref{wick}, it is straightforward to show $\normord{T^n}\, = \mathbf{He}_n(T|q)$, and the proof is given in Appendix \ref{proofT}. The Hermite polynomials of the transfer matrix satisfy
\begin{align}
\label{normcond}
     \bra{0}\mathbf{He}_m(T|q)\mathbf{He}_n(T|q)\ket{0} = \delta_{n,m}[n]_q!.
\end{align}
Defining the eigenstates of $T$ as $T\ket{E}\equiv E\ket{E}$, and the wavefunctions $\psi_n(E)\equiv \bra{E}\ket{n}$, 
\begin{align}
    \bra{E}\normord{T^n}\ket{0} = \bra{E}\ket{n}=\bra{E}\mathbf{He}_n(T|q){\ket{0}} = \mathbf{He}_n\left(E|q\right)\psi_0(E).
\end{align} 
The kets are normalized to unity, $\mathbb{I}=\int dE\,\ket{E}\bra{E}$, 
and $\psi_0(E)$ is fixed from the normalization condition Eq.~\eqref{normcond},
\begin{align}
  \bra{0}\mathbf{He}_m(T|q)\mathbf{He}_n(T|q)\ket{0} =  
  \int \, dE \ \abs{\psi_0(E)}^2 \, \mathbf{He}_m(E|q)\mathbf{He}_n(E|q) =[n]_q!\delta_{n,m} ,
\end{align} 
which shows that 
\begin{align}
    \abs{\psi_0(E)}^2=\rho(E|q),
\end{align}
where $\rho(E|q) $ is the $q$-normal distribution defined in Eq.~\eqref{qnormal}.
Therefore,
\begin{align}
    \bra{0}T^n\ket{0} = \int \, dE \ \abs{\psi_0(E)}^2  E^n =\langle H^n \rangle,
\end{align}
which is the duality between moments of DS-models and ground state expectation values of the transfer matrix. The transfer matrix can then be interpreted as the Hamiltonian in the dual Hilbert space.
\subsection{Operators and $n$-point functions}
Consider a second set of oscillators that act on the Hilbert space $\mathcal{H}_b$. The total Hilbert space is now $\mathcal{H}_{a\otimes b} = \mathcal{H}_a \otimes \mathcal{H}_b$. The second set of oscillators satisfy the following commutation relations\footnote{It is not necessary to define the commutator between $a^{\dagger}$ and $b^{\dagger}$.},
\begin{align}
    b b^{\dagger} - \rho_{12}b^{\dagger}b =1, \hspace{.5cm} a b^{\dagger} = \rho b^{\dagger}a,
\end{align}
We will show the dual of the operator $O$ considered in Sec.~\ref{strength} is 
\begin{align}
\label{O}
    O \to b^{\dagger}.
\end{align}
Another choice for the operator is $O \to b +b^{\dagger}$, which would make $O$ Hermitian, as in the SYK model.\footnote{The choice would effect the $4$-point function. For example, in SYK there are two uncrossed $4$-point functions, whereas the EGUE has one.}
\subsubsection{$2$-point function}
By considering the strength density,
\begin{align}
    \bra{0}b\,\delta(T-E_1)\,b^{\dagger}\,\delta(T-E_2)\,\ket{0},
\end{align}
the $2$-point function can be computed in the same way that was done in Eqs.~\eqref{2_point1}-\eqref{2point_2},
\begin{equation}
\begin{aligned}
     &O(E_1,E_2|\rho) =\sum_{n_1, n_2} \frac{\mathbf{He}_{n_1}(E_1|q)\mathbf{He}_{n_2}(E_2|q)}{[n_1]_q![n_2]_q!}\bra{0} b \,\normord{T^{n_1}} \,b^{\dagger} \,\normord{T^{n_2}} \ket{0} .
\end{aligned}
\end{equation}
The expectation value is straightforward to evaluate,
\begin{equation}
\begin{aligned}
    \bra{0} b \,\normord{T^{n_1}} \,b^{\dagger} \,\normord{T^{n_2}} \ket{0} &= \bra{0} b \,(a)^{n_1} \,b^{\dagger} \,(a^{\dagger})^{n_2} \ket{0} \\
    &=\rho^{n_1}\bra{0} b \,b^{\dagger} \,(a)^{n_1}  \,(a^{\dagger})^{n_2} \ket{0}\\
    &=\rho^{n_1}[n_1]_q! \delta_{n_1,n_2},
\end{aligned}
\end{equation}
so that the $2$-point function is the same as  Eq.~\eqref{2point_3},
\begin{align}
     O(E_1(\theta_1),E_2(\theta_2)|\rho) = \frac{(\rho^2)_{\infty}}{(\rho e^{i(\theta_1-\theta_2)},\rho e^{-i(\theta_1-\theta_2)},\rho e^{i(\theta_1+\theta_2)},\rho e^{-i(\theta_1+\theta_2)})_{\infty}}.
 \end{align}
 \subsubsection{$4$-point function}
 The $4$-point function is similarly computed by considering
 \begin{equation}
    \begin{aligned}
    \label{b4point}
 \bra{0}\,b\,\delta(T-E_1)\,b\,\delta(T-E_2)\,b^{\dagger}\,\delta(T-E_3)\,b^{\dagger}\,\delta(T-E_4)\ket{0}.
 \end{aligned}
 \end{equation}
 This will lead to two terms, corresponding to the uncrossed and crossed $4$-point functions. For purposes of presentation and matching the notation of Sec.~\ref{strength}, we introduce the oscillators $b_1$ and $b_2$, which satisfy $a b_i^{\dagger}=\rho_i b_i^{\dagger}a$ and $b_1 b_2^{\dagger} = \rho_{12} b_2^{\dagger} b_1$. Setting $\rho_1=\rho_2=\rho$ will recover the result of  Eq.~\eqref{b4point}.  To compute the uncrossed $4$-point function it is necessary to evaluate,
    \begin{equation}
    \begin{aligned}
    \label{uncrossedb}
\bra{0} b_1 \,\normord{T^{n_1}} \,b_2 \,\normord{T^{n_2}} b_2^{\dagger} \,\normord{T^{n_3}} \,b_1^{\dagger} \,\normord{T^{n_4}}  \ket{0} &= \bra{0} b_1 \,a^{n_1} \,b_2 \,\normord{T^{n_2}} \,b_2^{\dagger} \,\normord{T^{n_3}} \,b_1^{\dagger} \,(a^{\dagger})^{n_4}  \ket{0}.
\end{aligned} 
   \end{equation} 
   Note that,
   \begin{align}
   \label{identity1}
       \normord{T^n} \, = \sum_{k=0}^n \binom{n}{k}_q (a^{\dagger})^{n-k}a^{k} =\sum_{k=0}^n \binom{n}{k}_q (a^{\dagger})^{k}a^{n-k},
   \end{align}
   and 
   \begin{align}
   \label{identity2}
       a^{n}(a^{\dagger})^m = \sum_{k}\binom{n}{k}_q\binom{m}{k}_q[k]_q! (q^{m-k})^{n-k}(a^{\dagger})^{m-k}a^{n-k},
   \end{align}
   both of which can be proved by induction.  It then follows from  Eq.~\eqref{identity1},
   \begin{equation}
       \begin{aligned}
           b_2 \,\normord{T^{n_2}}\, b_2^{\dagger} &= \sum_k\binom{n_2}{k}_q \, b_2 \, (a^{\dagger})^{n_2-k}a^{k}\,b_2^{\dagger} =  \rho_2^{n_2} \sum_k\binom{n_2}{k}_q  \, (a^{\dagger})^{n_2-k}\,b_2 b_2^{\dagger}\,a^{k}\, = \rho_2^{n_2} \normord{T^{n_2}}.
       \end{aligned}
       \end{equation}
       where we used $b_2b_2^{\dagger} = 1 + \normord{b_2 b_2^{\dagger}}$ and dropped the normal ordered term. The expectation value Eq.~\eqref{uncrossedb} can be evaluated:
      \begin{equation}
    \begin{aligned}
    \label{uncrossedb1}
&\rho_2^{n_2} \sum_{k} \binom{n_2}{k}_q \bra{0}b_1 \,a^{n_1}  \, (a^{\dagger})^{k}a^{n_2-k} \,\normord{T^{n_3}} \,b_1^{\dagger} \,(a^{\dagger})^{n_4}  \ket{0}  \\
&=\rho_2^{n_2} \sum_{k,l} \binom{n_2}{k}_q \binom{n_1}{k}_q\binom{n_3}{l}_q[k]_q!\bra{0}b_1 \,a^{n_1+n_2-2k} \,(a^{\dagger})^{n_3-l}a^l\,b_1^{\dagger} \,(a^{\dagger})^{n_4}  \ket{0} \\
& =\rho_2^{n_2} \sum_{k,l} \binom{n_2}{k}_q \binom{n_1}{k}_q\binom{n_3}{l}_q\binom{n_4}{l}_q\binom{n_1+n_2-2k}{n_3-l}_q[n_3-l]_q![l]_q![k]_q! \rho_1^{n_1+n_2-2k-n_3+2l} \\
&\times \bra{n_1+n_2-2k -n_3+l}\ket{n_4-l} \\
& = \rho_2^{n_2}\rho_1^{n_4} \sum_{k,l} \binom{n_2}{k}_q \binom{n_1}{k}_q\binom{n_3}{l}_q\binom{n_4}{l}_q[n_1+n_2-2k]_q![l]_q![k]_q! \delta_{n_1+n_2-2k,n_3+n_4-2l}.
\end{aligned} 
   \end{equation} 
   To get this result we made repeated use of the identities Eqs.~\eqref{identity1} and~\eqref{identity2}. The result,  Eq.~\eqref{uncrossedb1}, is equivalent to  Eq.~\eqref{uncrossedO} which proves the uncrossed $4$-point function is equivalent to  Eq.~\eqref{uncrossed_final}. For the crossed $4$-point function it is necessary to evaluate the expectation value 
      \begin{equation}
    \begin{aligned}
    \label{crossedb}
\bra{0} b_1 \,\normord{T^{n_1}} \,b_2 \,\normord{T^{n_2}} b_1^{\dagger} \,\normord{T^{n_3}} \,b_2^{\dagger} \,\normord{T^{n_4}}  \ket{0} .
\end{aligned} 
   \end{equation} 
   The identities, Eqs.~\eqref{identity1} and~\eqref{identity2}, along with similar steps as  Eq.~\eqref{uncrossedb1}, can be used to show  Eq.~\eqref{crossedb} is equivalent to  Eq.~\eqref{crossedO}, which proves the $4$-point function is equivalent to  Eq.~\eqref{crossedOfinal}. 

   Therefore, there is an exact duality between $(O,H, \mathcal{H}_{\text{EGUE}})$ and $(b^{\dagger}, T, \mathcal{H}_{a\otimes b})$ in the double-scaled limit. In DS-SYK the duality is between $(O,H, \mathcal{H}_{\text{SYK}})$ and $(T_b, T_a, \mathcal{H}_{a\otimes b})$, with $T_b = b+b^{\dagger}$ and $T_a=a+a^{\dagger}.$
\section{Double-scaled complex SYK at fixed charge}
\label{complex}
The complex SYK  (cSYK)~\cite{Berkooz:2020, Davison_2017, Sachdev:2015efa, Gu2020} is just the EGUE, however, it is typically defined in a $2^N$ dimensional Hilbert space instead of a fixed-particle Hilbert space. Not working in a fixed-particle Hilbert space leads to unnecessary complications, since the moment method of Sec.~\ref{1} cannot be directly applied. Moreover, from the quantum chaos perspective it is necessary to account for all conserved quantities to see signatures of chaos \cite{Bhattacharya_2017}. That being said, in \cite{Berkooz:2020} the moments of DS-cSYK with fermions were derived through a careful analysis of chord diagrams. In this section, we will compare the results for the moments in~\cite{Berkooz:2020} to the results derived in this paper for the fermionic EGUE and explain the differences.

The Hamiltonian of cSYK defined in \cite{Berkooz:2020} is the same as the fermionic EGUE Eq.~\eqref{fegue}\footnote{The fermions in \cite{Berkooz:2020} were defined with an extra factor of $\sqrt{2}$ and we include a factor of $2^{-p}$ to account for this. We will include this factor in the moments as well.},
\begin{equation}
H_{\text{cSYK}} = 2^{-p}\sum_{I,L}J_{I,L}\psi^{\dagger}_{p,I}\psi_{p,L},
\end{equation}
with an important difference being the variance of $J_{I,L}$ normalized as
\begin{align}
\label{syknorm}
     \expval{J_{I,L}J_{I',L'}}_J =  \binom{N}{p}^{-2}\delta_{I,L'}\delta_{L,I'}.
\end{align}
The charge and charge density are defined by
\begin{align}
    Q \equiv m - \frac{N}{2}, \hspace{.5cm} \mathcal{Q} \equiv \frac{Q}{N}.
\end{align}
In \cite{Berkooz:2020}, the moments of the grand-canonical ensemble were considered
\begin{align}
  m_{2k}(\mu)\equiv  \ev{\tr H_{\text{cSYK}}^{2k}e^{-2\mu Q}}_J,
\end{align}
where the trace is in the $2^N$ dimensional space.
We refer the readers to \cite{Berkooz:2020} for details of the derivation and the exact expression. The result was projected onto a fixed-particle space by performing a contour integration over $\mu$ in a large $N$ limit. To be concrete, defining $z = e^{2\mu}$, they showed
\begin{equation}
\begin{aligned}
\label{sykmoments}
   \oint dz \frac{m_{2k}(z)}{z^{-Q+1}}&\approx  \frac{2}{\sqrt{2\pi N}}\frac{1}{\sqrt{1-4\mathcal{Q}}}\exp\left\{N\mathcal{Q}\log(\frac{1-2\mathcal{Q}}{1+2\mathcal{Q}})+\frac{N}{2}\log(\frac{4}{1-4\mathcal{Q}^2})\right\} \\
   &\times \int \,dE \, \rho(E,q) E^{2k}\left(4^{-p}\left(1-4\mathcal{Q}^2\right)^{p}\exp\left\{p^2/N + \frac{4p^2 \mathcal{Q}}{N(1-4\mathcal{Q}^2)}\right\}\right)^{k}.
\end{aligned}
\end{equation}
The prefactor, as was explained in \cite{Berkooz:2020}, comes from not normalizing the trace in the fixed particle sector. This gives a factor of $\binom{N}{m}$, which after using Sterlings approximation becomes
\begin{align}
    \binom{N}{m}=\binom{N}{Q+\frac{N}{2}} \sim \frac{2}{\sqrt{2\pi N}}\frac{1}{\sqrt{1-4\mathcal{Q}}}\exp\left\{NQ\log(\frac{1-2Q}{1+2Q})+\frac{N}{2}\log(\frac{4}{1-4\mathcal{Q}^2})\right\},
\end{align}
which is equivalent to the prefactor in Eq.~\eqref{sykmoments}.
The scaling factor of the energy is explained by the choice of variance in Eq.~\eqref{syknorm} compared to Eq.~\eqref{eguenorm},
\begin{equation}
\begin{aligned}
    \frac{\binom{m}{p}\binom{N-m+p}{p}}{\binom{N}{p}\binom{N}{p}} &\sim \left(\frac{m}{N}-\frac{m^2}{N^2}\right)^{p} \exp\left\{p^2/N  +\frac{p^2(\frac{m}{N}-1/2)}{m(1-m/N)}\right\} \\
    &=4^{-p}\left(1-4\mathcal{Q}^2\right)^{p}\exp\left\{p^2/N + \frac{4p^2 \mathcal{Q}}{N(1-4\mathcal{Q}^2)}\right\}.
\end{aligned}
\end{equation}
Therefore, the moments in  Eq.~\eqref{sykmoments} are not the reduced moments, which is why there are additional factors. The variance,  Eq.~\eqref{syknorm}, was specifically chosen to allow the chord diagram approach in the $2^N$ dimensional space, and it is not obvious that the simplest choice of variance is actually  Eq.~\eqref{eguenorm} unless one does the computation,  Eq.~\eqref{2nmoment}, in the fixed-particle space. The $2$- and $4$-point functions for fermionic DS-cSYK were also computed in \cite{Berkooz:2020} with chord techniques, and the additional factors that appear can be explained in a similar manner.

We would like to emphasize the simplicity of evaluating the moments in the fixed-particle space using the techniques of Sec.~\ref{1}, compared to the steps necessary in deriving  Eq.~\eqref{sykmoments}. Furthermore, the methods of Sec.~\ref{1} can be directly generalized to bosons, whereas it is not obvious if the same is true for the chord diagram analysis of \cite{Berkooz:2020}.
\section{Conclusion}
\label{conclusion}
Embedded ensembles are very old models of many-body quantum chaos first studied analytically more than 50 years ago \cite{Mon1975}. We have shown in the double-scaled limit embedded ensembles have the exact same properties as the SYK model. This broadens the double-scaled universality class to include both fermionic and bosonic systems. We worked specifically with the EGUE, but in the double-scaled limit the results are also valid for the time-reversal invariant case, i.e. the EGOE and EGSE.

Embedded ensembles are known to be solved, beyond the double-scaled limit, in terms of the Wigner Racah coefficients of $U(N)$ \cite{Kota2014}. Developing an analogous formalism for other $p$-body systems, e.g. SYK, could provide a group theoretical understanding of the double-scaled universality class. In the double-scaled limit there are connections to the quantum group $SU_q(1,1)$ \cite{Berkooz:2018jqr, Berkooz:2024}. For example, the $4$-point function is related to Wigner $6j$-symbol of this group \cite{Berkooz:2018jqr, Berkooz:2024}. It is natural to ask if there are any connections between the $U(N)$ solution valid at any $N$ and the $SU_q(1,1)$ solution valid in the double-scaled limit. This might be connected to the work of \cite{Lin2023}.

Traditional random matrix models admit a pertubative genus expansion and there are efficient analytic techniques to compute correlation functions to all orders in this expansion \cite{Eynard2018}. The genus expansion is essential for understanding the connection to two-dimensional gravity, since the gravitational path integral has the same expansion \cite{Saad2019}. For traditional matrix models the genus expansion is suppressed by powers of the dimensionality. Embedded ensembles become the GUE when $p=m$, and the genus expansion is suppressed by $\binom{N}{m}$. It was found in \cite{Brody1981} that fluctuations, i.e. the two-trace correlator, for $m\gg p$ are suppressed by powers of $\binom{m}{p}$ for the EGOE.  Assessing these regimes in $p$-body systems is difficult using the moment method since $1/N$ corrections are neglected. Different approaches have been taken, such as the diagrammatic method based on chords developed in \cite{Berkooz:2020fvm}, or the sigma model approach of \cite{Altland:2017eao}. The approach of this paper, based on normal mode expansions similar to \cite{Brody1981}, could be useful in this regard. The previously mentioned group theoretical techniques based on the $U(N)$ Wigner Racah algebra could also play an important role. Progress for embedded ensembles in this regard has been made in \cite{Kota2023}.

We have developed new techniques to compute the density of states and $n$-point functions, and these techniques allow the $n$-point functions to be computed directly in the energy basis. For the $2$-point function, or strength density, the formalism makes it very easy to take the limiting cases of $q\to 1$ and $q \to 0$. It would be interesting to compute the same limits for the $4$-point function. The limit $q\to 1$ is typically taken while keeping low temperature corrections \cite{Baur_2023, Goel:2023svz} due to its importance for AdS$_2$ holography. However, at the center of the spectrum the results are still physically relevant for many-body chaos, and the results simplify dramatically. For example, the strength density at the center of the spectrum is just a bivariate Gaussian when $q \to 1$. Understanding the $n$-point strength density as $q \to 1$ and $q \to 0$ has important implications for the role of few-body interactions in generalized ETH.

\section*{Acknowledgments}
We would like to thank Torsten Weber, Juan Diego Urbina, and Klaus Richter for useful discussions. We would also like to thank the hospitality of the Institut f\"ur Theoretische Physik at the
Universit\"at Regensburg where this collaboration initially began.


\begin{appendix}

\section{$q$-formulas and the $4$-point function}
\label{appendixa}
 In this section we collect some useful $q$-formulas and give the details of evaluating the sums that occurs in the $4$-point function.
 \subsection{q-formulas}
 \label{qappendix}
 The $q$-Pochhamer symbol is defined as
 \begin{align}
     (a)_n \equiv (a;q)_n \equiv \prod_{k=0}^{n-1}(1-aq^{k})
 \end{align}
 and 
 \begin{align}
     (a_1,a_2,\dots,a_n)_\infty =(a_1)_\infty (a_2)_\infty \dots (a_n)_\infty .
 \end{align}
 The $q$-factorial is defined as
 \begin{equation}
     [n]_q! = \frac{(q)_n}{(1-q)^n}
 \end{equation}
 and the $q$-binomial coefficient
 \begin{equation}
     \binom{n}{k}_q = \frac{[n]_q!}{[n-k]_q![k]_q!}.
 \end{equation}
The $q$-binomial theorem is
\begin{align}
\label{q_binom}
    (z)_{n}= \sum_k \binom{n}{k}_q  q^{\binom{k}{2}} (-z)^k
\end{align}
and the $q$-Vandermonde identity
\begin{align}
    \binom{n+m}{k}_q = \sum_{s_1+s_2=k}\binom{n}{s_1}_q\binom{m}{s_2}_q q^{\binom{k}{2}}q^{s_2(n-s_1)}.
\end{align}
We also define
\begin{align}
     O(E_1,E_2|\rho) &= \sum_{k}\frac{\rho^k\mathbf{He}_{k}(E_1|q)\mathbf{He}_{k}(E_2|q)}{[k]_q!} \\
     &=(\rho e^{i(\theta_1-\theta_2)},\rho e^{-i(\theta_1-\theta_2)},\rho e^{i(\theta_1+\theta_2)},\rho e^{-i(\theta_1+\theta_2)})_{\infty}
 \end{align}
 with $E_i=\frac{2\cos(\theta_i)}{\sqrt{1-q}}$. 
Furthermore,
\begin{equation}
\label{sumH}
    O(E_1,E_2|\rho)Q_{n,m}(E_1,E_2|\rho, q) = \sum_{k}\frac{\rho^k\mathbf{He}_{n+k}(E_1|q)\mathbf{He}_{m+k}(E_2|q)}{[k]_q!}
\end{equation}
where \cite{szabowski2013}
\begin{equation}
  Q_{n,m}(E_1,E_2|\rho, q) = \sum_{s}\binom{m}{s}_q(-1)^s q^{\binom{s}{2}} \rho^s \mathbf{P}_{n+s}(E_1|E_2,\rho,q)\mathbf{He}_{m-s}(E_2|q)/(\rho^2)_{n+s}.
\end{equation}
The polynomials $\mathbf{P}_{n}(E_1|E_2,\rho,q)$ are closely related to the Al-Salam-Chihara polynomials
\begin{equation}
    \mathbf{P}_{n}(E_1|E_2,\rho,q) = \frac{1}{(1-q)^{n/2}}\mathbf{Q}_n\left(\cos(\theta_1)|\rho e^{i\theta_2}, \rho e^{-i\theta_2},q\right),
\end{equation}
and they satisfy the identity \cite{askey1994,szabowski2013},
\begin{align}
\label{pident}
   & \sum_{n}\frac{ \rho_1^n\mathbf{P}_{n}(E_1|E_2,\rho_2,q) \mathbf{P}_{n}(E_4|E_3,\rho_2,q)}{[n]_q!(\rho_2^2)_n} =\\[8pt]
   &\frac{(\rho_1e^{-i(\theta_2+\theta_3)},\rho_2 \rho_1 e^{i(\theta_3\pm \theta_1)},\rho_2 \rho_1 e^{i(\theta_2\pm \theta_4)})_{\infty}}{(\rho_1 \rho_2^2 e^{i(\theta_2+\theta_3)},\rho_1^2)_{\infty}}O(E_1,E_4|\rho_1) \\
   &{}_8W_7\left(q^{-1}\rho_1\rho_2^2 e^{i(\theta_2+\theta_3)};\rho_1e^{i(\theta_3+\theta_2)},\rho_2e^{i(\theta_2 \pm \theta_1)},\rho_2e^{i(\theta_3 \pm \theta_4)};q,\rho_1 e^{-i(\theta_3+\theta_2)}\right)
   \end{align}
   where the $\pm$ notations means taking both signs, e.g. $(\rho_2 \rho_1 e^{i(\theta_3\pm \theta_1)}) = (\rho_2 \rho_1 e^{i(\theta_3+ \theta_1)},\rho_2 \rho_1 e^{i(\theta_3- \theta_1)}) $. The ${}_8W_7$ is a basic hypergeometric series is defined as \cite{Berkooz:2018jqr}
   \begin{align}
   \label{Wdef}
       {}_8W_7(a;b,c,d,e,f;q,z) \equiv \sum_{n=0}^{\infty}\frac{(a,\pm q a^{1/2},b,c,d,e,f)_n}{(\pm a^{1/2},qa/b,qa/c,qa/d,qa/e,qa/f,q)_n}z^n.
   \end{align}
\subsection{Uncrossed $4$-point function}
\label{uncrossed_appendix}
The sum appearing in the uncrossed $4$-point function, Eq.~\eqref{4point2}, can now be evaluated,
\begin{equation}
    \sum_{n_1, n_2,n_3,n_4,k,l} \frac{\rho_{1}^{n_4+l}\rho_2^{n_2+k}\mathbf{He}_{n_1+k}(E_1|q)\mathbf{He}_{n_2+k}(E_2|q)\mathbf{He}_{n_3+l}(E_3|q)\mathbf{He}_{n_4+l}(E_4|q)[n_1+n_2]_q!}{[n_1]_q![n_2]_q![n_3]_q![n_4]_q! [k]_q![l]_q!}\delta_{n_1+n_2,n_3+n_4}.
\end{equation}
Summing over $k$ and $l$ gives
\begin{equation}
\begin{aligned}
   &O(E_1,E_2|\rho_2)O(E_3,E_4|\rho_1) \sum_{n_1, n_2}\frac{\rho_2^{n_2}Q_{n_1,n_2}(E_1,E_2|\rho_2,q)}{[n_1]_q![n_2]_q!} \sum_{n_3}\rho_1^{n_1+n_2-n_3}Q_{n_3,n_1+n_2-n_3}(E_3,E_4|\rho_1,q)\binom{n_1+n_2}{n_3}_q,
   \end{aligned}
\end{equation}
and the following identity can be used \cite{Szabowski2011},
\begin{align}
    \mathbf{He}_{n_1+n_2}(E_3|q)=\sum_{n_3}\rho_1^{n_1+n_2-n_3}Q_{n_3,n_1+n_2-n_3}(E_3,E_4|\rho_1,q)\binom{n_1+n_2}{n_3}_q.
\end{align}
The remaining sums over $n_1$ and $n_2$ are then
\begin{equation}
    \begin{aligned}
&\sum_{n_1,n_2}\frac{\rho_2^{n_2}Q_{n_1,n_2}(E_1,E_2|\rho_2,q)\mathbf{He}_{n_1+n_2}(E_4|q)}{[n_1]_q![n_2]_q!} =\\
        & \sum_{n_1,n_2,s}\frac{\mathbf{He}_{n_1+n_2}(E_4|q)\rho_2^{n_2}}{[n_1]_q![n_2]_q!} \binom{n_2}{s}_q(-1)^s q^{\binom{s}{2}} \rho^s \mathbf{P}_{n_1+s}(E_1|E_2,\rho_2,q)\mathbf{He}_{n_2-s}(E_2|q)/(\rho_2^2)_{n_1+s},
    \end{aligned}
\end{equation}
substituting $n_2 \to n_2 + s$ and $n_1 \to n_1 - s$,
\begin{equation}
\begin{aligned}
&\sum_{n_1,n_2}\frac{\mathbf{He}_{n_1+n_2}(E_3|q)\rho_2^{n_2}}{[n_1]_q![n_2]_q!}\mathbf{P}_{n_1}(E_1|E_2,\rho_2,q)\mathbf{He}_{n_2}(E_2|q)/(\rho_2^2)_{n_1}\sum_{s} \binom{n_1}{s}_q(-1)^s q^{\binom{s}{2}} \rho_2^{2s}  \\
&=\sum_{n_1,n_2}\frac{\rho_2^{n_2}}{[n_1]_q![n_2]_q!}\mathbf{He}_{n_1+n_2}(E_3|q)\mathbf{P}_{n_1}(E_1|E_2,\rho_2,q)\mathbf{He}_{n_2}(E_2|q) \\
& = O(E_2,E_3|\rho_2)\sum_{n_1}\frac{1}{[n_1]_q!(\rho_2^2)_{n_1}}\mathbf{P}_{n_1}(E_3|E_2,\rho_2,q)\mathbf{P}_{n_1}(E_1|E_2,\rho_2,q) \\
& =\frac{O(E_2,E_3|\rho_2)\delta(E_1-E_3)}{O(E_1,E_2|\rho_2)\rho(E_1|q)}. 
\end{aligned}
\end{equation}
To get the second line we used the $q$-binomial theorem Eq.~\eqref{q_binom}, to get the third line we used Eq.~\eqref{sumH}, and the last line follows from the orthogonality of the polynomials $\mathbf{P}_{n_1}(E_1|E_2,\rho_2,q)$ \cite{szabowski2013}. The uncrossed $4$-point function is then,
\begin{align}
    O_{\text{uncrossed}}(E_1,E_2,E_3,E_4|\rho_1,\rho_2) = O(E_3,E_4|\rho_1)O(E_2,E_3|\rho_2)\frac{\delta(E_1-E_3)}{\rho(E_1|q)}.
\end{align}
\subsection{Crossed $4$-point function}
\label{crossed_appendix}
In this section we evaluate the sum,
 \begin{equation}
    \begin{aligned}
        &\rho_{12}^{-1}O_{crossed}(E_1,E_2,E_3,E_4|\rho_1,\rho_2)=\\[8pt]
        &\sum_{n_1, n_2,n_3,k,l,j} \frac{\mathbf{He}_{n_1+k+j}(E_1|q)\mathbf{He}_{n_2+k+l}(E_2|q)\mathbf{He}_{n_3+j+l}(E_3|q)\mathbf{He}_{n_1+n_2+n_3}(E_4|q)}{[n_1]_q![n_2]_q![n_3]_q![j]_q![k]_q![l]_q!} \\
        &\times \rho_1^{n_1+n_2+l+j}\rho_2^{n_2+n_3+k+j}q^{jn_2}.
        \end{aligned}
    \end{equation}
Summing over $k$ and $n_3$ will give a factor of $O(E_1,E_2|\rho_2)O(E_3,E_4|\rho_2)$ multiplied by:
\begin{equation}
    \begin{aligned}
&\sum_{n_1,n_2,l,j}\frac{Q_{n_1+j,n_2+l}(E_1,E_2|\rho_2,q)Q_{n_1+n_2,j+l}(E_4,E_3|\rho_2,q)\times \rho_1^{n_1+n_2+l+j}\rho_2^{n_2+j}q^{jn_2}}{[n_1]_q![n_2]_q![j]_q![l]_q!}=\\[8pt]
        &\sum_{n_1,n_2,l,j}\frac{Q_{n_1+n_2,j+l}(E_4,E_3|\rho_2,q)\times \rho_1^{n_1+n_2+l+j}\rho_2^{n_2+j}q^{jn_2}}{[n_1]_q![n_2]_q![j]_q![l]_q!} \\
        &\times \sum_{s}\binom{n_2+l}{s}_q(-1)^s q^{\binom{s}{2}} \rho_2^s \mathbf{P}_{n_1+j+s}(E_1|E_2,\rho_2,q)\mathbf{He}_{n_2+l-s}(E_2|q)/(\rho_2^2)_{n_1+j+s} =\\
           &\sum_{n_1,n_2,l,j}\frac{Q_{n_1+n_2,j+l}(E_4,E_3|\rho_2,q)\times \rho_1^{n_1+n_2+l+j}\rho_2^{n_2+j}q^{jn_2}}{[n_1]_q![n_2]_q![j]_q![l]_q!} \\
        &\times \sum_{s_1,s_2}\binom{n_2}{s_1}_q\binom{l}{s_2}_q q^{s_2(n_2-s_1)}(-1)^{s} q^{\binom{s_1}{2}+\binom{s_2}{2}+s_1s_2} \rho_2^{s} \mathbf{P}_{n_1+j+s}(E_1|E_2,\rho_2,q)\mathbf{He}_{n_2+l-s}(E_2|q)/(\rho_2^2)_{n_1+j+s}
    \end{aligned}
\end{equation}
where in the last line we used the $q$-Vandermonde identity with $s=s_1+s_2$. Substituting $n_2\to n_2+s_1$, $l\to l_2+s_2$, $n_1 \to n_1-s_1$, $j \to j - s_2$, gives
\begin{equation}
    \begin{aligned}
        &\sum_{n_1,n_2,l,j}\frac{Q_{n_1+n_2,j+l}(E_4,E_3|\rho_2,q)\times \rho_1^{n_1+n_2+l+j}\rho_2^{n_2+j}q^{jn_2}}{[n_1]_q![n_2]_q![j]_q![l]_q!}\mathbf{P}_{n_1+j}(E_1|E_2,\rho_2,q)\mathbf{He}_{n_2+l}(E_2|q)/(\rho_2^2)_{n_1+j} \\
        &\times \sum_{s_1,s_2}(-1)^{s_1+s_2} q^{\binom{s_1}{2}+\binom{s_2}{2}} \rho_2^{2s_1}q^{s_1j}\binom{n_1}{s_1}_q\binom{j}{s_2}_q = \\[8pt]
        & \sum_{n_1,n_2,l,j}\frac{Q_{n_1+n_2,j+l}(E_4,E_3|\rho_2,q)\times \rho_1^{n_1+n_2+l+j}\rho_2^{n_2+j}q^{jn_2}}{[n_1]_q![n_2]_q![j]_q![l]_q!}\mathbf{P}_{n_1+j}(E_1|E_2,\rho_2,q)\mathbf{He}_{n_2+l}(E_2|q)/(\rho_2^2)_{n_1+j} \\
        &\times (\rho_2^2q^j)_{n_1}(1)_j
    \end{aligned}.
\end{equation}
Note that $(1)_j=\delta_{j,0}$ so this becomes
\begin{equation}
    \begin{aligned}
        \sum_{n_1,n_2,l}\frac{Q_{n_1+n_2,l}(E_4,E_3|\rho_2,q)\times \rho_1^{n_1+n_2+l}\rho_2^{n_2}}{[n_1]_q![n_2]_q![l]_q!}\mathbf{P}_{n_1}(E_1|E_2,\rho_2,q)\mathbf{He}_{n_2+l}(E_2|q) .
    \end{aligned}
\end{equation}
Essentially following the same steps with $Q_{n_1+n_2,l}$ leads to 
\begin{equation}
    \begin{aligned}
    &\sum_{n_1,l} \frac{\mathbf{P}_{n_1}(E_1|E_2,\rho_2,q)\mathbf{P}_{n_1}(E_4|E_3,\rho_2,q)\mathbf{He}_{l}(E_2|q)\mathbf{He}_{l}(E_3|q)\rho_1^{n_1+l}}{[n_1]_q![l]_q!} =\\
    &O(E_2,E_3|\rho_1) O(E_1,E_4|\rho_1)\frac{(\rho_1e^{-i(\theta_2+\theta_3)},\rho_2 \rho_1 e^{i(\theta_3\pm \theta_1)},\rho_2 \rho_1 e^{i(\theta_2\pm \theta_4)})_{\infty}}{(\rho_1 \rho_2^2 e^{i(\theta_2+\theta_3)},\rho_1^2)_{\infty}} \\
   &{}_8W_7\left(q^{-1}\rho_1\rho_2^2 e^{i(\theta_2+\theta_3)};\rho_1e^{i(\theta_3+\theta_2)},\rho_2e^{i(\theta_2 \pm \theta_1)},\rho_2e^{i(\theta_3 \pm \theta_4)};q,\rho_1 e^{-i(\theta_3+\theta_2)}\right)
    \end{aligned}
\end{equation}
where we used Eq.~\eqref{pident}. The crossed $4$-point function is then
\begin{equation}
    \begin{aligned}
        &O_{crossed}(E_1,E_2,E_3,E_4|\rho_1,\rho_2) = \rho_{12} \times \\[8pt]
         &O(E_1,E_2|\rho_2)O(E_3,E_4|\rho_2)O(E_2,E_3|\rho_1) O(E_1,E_4|\rho_1)\frac{(\rho_1e^{-i(\theta_2+\theta_3)},\rho_2 \rho_1 e^{i(\theta_3\pm \theta_1)},\rho_2 \rho_1 e^{i(\theta_2\pm \theta_4)})_{\infty}}{(\rho_1 \rho_2^2 e^{i(\theta_2+\theta_3)},\rho_1^2)_{\infty}} \\
   &{}_8W_7\left(q^{-1}\rho_1\rho_2^2 e^{i(\theta_2+\theta_3)};\rho_1e^{i(\theta_3+\theta_2)},\rho_2e^{i(\theta_2 \pm \theta_1)},\rho_2e^{i(\theta_3 \pm \theta_4)};q,\rho_1 e^{-i(\theta_3+\theta_2)}\right).
    \end{aligned}
    \end{equation}
\section{Review of chords and the transfer matrix}
\label{review}
\begin{figure}
      \centering
      \includegraphics[width=0.75\linewidth]{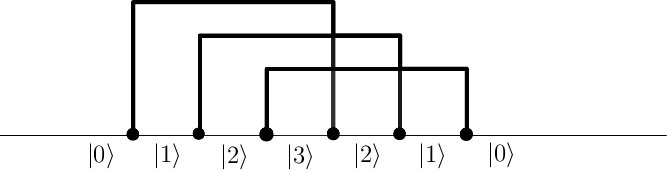}
      \caption{Example of counting open chords for the $ABCACB$ diagram term in the 6th moment.}
      \label{fig:6thmoment}
  \end{figure}

\begin{figure}
      \centering
      \includegraphics[width=.75\linewidth]{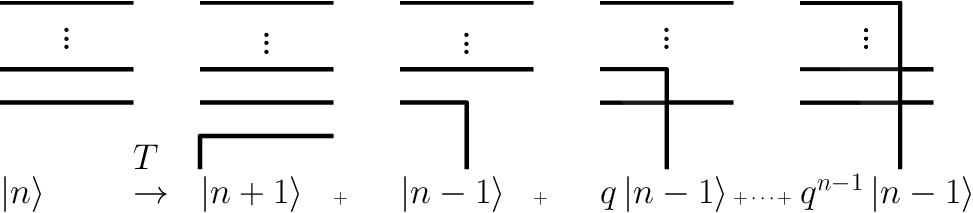}
      \caption{Application of the transfer matrix on a state of $n$ open chords.}
      \label{fig:transfer}
  \end{figure}

In this section we review the chord diagram approach to deriving the density of states developed in \cite{Berkooz:2018qkz, Berkooz:2018jqr, Berkooz:2024}. Moments of double-scaled models can be written as a sum over chord diagrams,
  \begin{align}
      \expval{H^{2k}} = \sum_{\text{chord diagrams w/
     $k$ chords}} \left(q\right)^{\text{\# of intersections}}.
  \end{align}
We first introduce the chord Hilbert space which is defined by the space of open chords. For example, the state with $n$ open chords is represented by the basis vector,
  \begin{align}
      \ket{n} =\ket{\underbrace{0,\dots,0}_{n-1},1,0,\dots}.
  \end{align}
  This is taken to be an infinite dimensional Hilbert space, i.e. 
   \begin{align}
     \mathcal{H} = \Big \{ \,\ket{n}\,  | \, n \in \mathbb{Z_+}\,\Big\}.
 \end{align}
  An example of counting open chords is given for the $ABCACB$ diagram of the sixth moment in fig.~\ref{fig:6thmoment}.
  The key point is that for an arbitrary chord diagram, at a state of $n$ open chords, there are only two options that can occur: At the next node a chord can open resulting in a state of $n+1$ open chords, or anyone of the $n$ open chords can close resulting in a state of $n-1$ open chords. This is depicted in fig.~\ref{fig:transfer}. The operator that evolves the state from one node to the next is refereed to as the transfer matrix $T$.
 We can then write the operation of the transfer matrix on a state $\ket{n}$ as
  \begin{align}
  \label{recursion}
      T\ket{n}=\ket{n+1}+\sum_{i=0}^{n-1}q^i \ket{n-1} = \ket{n+1}+[n]_q \ket{n-1}.
  \end{align}
  Every diagram involved in the computation of $\expval{H^{2k}}$ begins with $0$ open chords, i.e. the ground state $\ket{0}$, and ends with $0$ open chords. Each diagram has $2k$ nodes and thus the transfer matrix needs to be applied $2k$ times. The solution for the moments is then
  \begin{align}
  \label{moments1}
      \expval{H^{2k}} = \bra{0}T^{2k}\ket{0}.
  \end{align}
  The transfer matrix can be written in terms of creation and annihilation operators of $q$-oscillators. The operators are defined as follows:
\begin{align}
    a\ket{n}=[n]_q \ket{n-1}, \hspace{.1cm} a^{\dagger}\ket{n}=\ket{n+1},
\end{align}
and they satisfy the $q$-commutator relation,
    \begin{align}
        [a,a^{\dagger}]_q=a a^{\dagger} - qa^{\dagger}a=1.
    \end{align}
   The norm on the Hilbert space is defined so that $a$ and $a^{\dagger}$ are hermitian conjugates,
   \begin{align}
   \label{ortho1}
       \bra{m}\ket{n}=[n]_q!\delta_{n,m}.
   \end{align}
   The transfer matrix can then be written as
   \begin{align}
       T = a+a^{\dagger}.
   \end{align}
The eigenstates and eigenvalues of $T$ will be denoted as $T\ket{E}=E\ket{E}$. The wavefunctions, $\psi_n(E)=\braket{E}{n}$, can be found using Eq.~\eqref{recursion},
\begin{align}
    \bra{E}T\ket{n} = E\psi_n(E) = \psi_{n+1}(E) + [n]_q \psi_{n-1}(E)
\end{align}
   From the recursion, it is seen the wave functions are proportional to the $q$-Hermite polynomials, which satisfy
   \begin{align}
    \mathbf{He}_{n+1}(E|q) = E \,\mathbf{He}_{n}(E|q) - [n]_q\mathbf{He}_{n-1}(E|q).
\end{align}
with $\mathbf{He}_{0}(E|q)=1$. This gives
  \begin{align}
  \label{wavefunction}
     \psi_n(E) = \psi_0(E)\mathbf{He}_n\left(E|q\right).
  \end{align}
  To solve for $\psi_0(E)$, we can insert a complete set of states, $\mathbb{I}=\int dE \ket{E}\bra{E}$, into the orthogonality condition Eq.~\eqref{ortho},
  \begin{align}
   \bra{m}\ket{n}=   \int \, dE \ \abs{\psi_0(E)}^2 \, \mathbf{He}_m(E|q)\mathbf{He}_n(E|q) =[n]_q!\delta_{n,m},
  \end{align}
  which show $|\psi_0(E)|^2$ must be the $q$-normal distribution defined in Eq.~\eqref{qnormal}. The moments are thus
 \begin{align}
    \bra{0}T^{2k}\ket{0} = \int \, dE \ \abs{\psi_0(E)}^2  E^{2k} =\langle H^{2k} \rangle.
\end{align}
The chord diagram method can be extended to compute $n$-point functions and we refer the reader to \cite{Berkooz:2018qkz, Berkooz:2018jqr, Berkooz:2024} for details.
\section{Normal ordering of the transfer matrix}
\label{proofT}
Here we prove that $\normord{T^n} \,\,=\mathbf{He}_n(T|q)$.  Start with $T \normord{T^n}$, which is a polynomial of degree $n+1$ in $\normord{T}$,
\begin{align}
    \label{def2}
    T\normord{T^n}\,\,= \sum_{j=0}^{n+1}b_j \normord{T^j}.
\end{align}
The definition Eq.~\eqref{normal_def} for $m< n-1$ shows $b_j =0$ when $ j < n-1$. Applying Eq.~\eqref{normal_def} for $m=n-1$ shows $b_{n-1} = [n]_q $, and for $m=n+1$,
\begin{align}
    \bra{0}T^{n+2}\normord{T^n}\ket{0} = b_{n+1}[n+1]_q! + [n]_q \bra{0}T^{n+1}\normord{T^{n-1}}\ket{0}.
\end{align}
To solve $\bra{0}T^{n+2}\normord{T^n}\ket{0}$, note that only $(a^{\dagger})^n$ in $\normord{T^n}$ can contribute so that there has to be a corresponding $a^{n}$ from $T^{n+2}$. We can choose a $T^2$ from $T^{n+2}$ to factor out, which will contribute  $(a^{2}+(a^{\dagger})^2 +aa^{\dagger} + a^{\dagger}a)$, but only the term $a a^{\dagger}$ can give a non-zero contribution. For example, if the right most $T^2$ is chosen,
\begin{align}
     \bra{0}T^{n}T^2 \to \bra{0}a^{n}(a a^{\dagger}) = [n+1]_q\bra{n}
\end{align}
Therefore,
\begin{align}
      \bra{0}T^{n+2}\normord{T^n}\ket{0} = [n]_q!\sum_{k=1}^{n+1} [k]_q, 
\end{align}
and 
\begin{align}
    b_{n+1} = \frac{[n]_q!}{[n+1]_q!} (\sum_{k=1}^{n+1}[k]_q-\sum_{k=1}^{n}[k]_q) = 1 .
\end{align}
The normal ordered product $\normord{T^n}$ then satisfies the recursion
\begin{align}
    \normord{T^{n+1}} \hspace{.1cm}= T\normord{T^n} \hspace{.1cm}-\hspace{.1cm} [n]_q \normord{T^{n-1}}
\end{align}
with $\normord{T^0} \hspace{.1cm}=1$, which proves $\normord{T^n}\hspace{.1cm}=\mathbf{He}_n(T|q)$.

\end{appendix}

\bibliography{lib.bib}

\begin{thebibliography}{10}
\providecommand{\url}[1]{\texttt{#1}}
\providecommand{\urlprefix}{URL }
\expandafter\ifx\csname urlstyle\endcsname\relax
  \providecommand{\doi}[1]{doi:\discretionary{}{}{}#1}\else
  \providecommand{\doi}{doi:\discretionary{}{}{}\begingroup \urlstyle{rm}\Url}\fi
\providecommand{\eprint}[2][]{\url{#2}}

\bibitem{Mehta2004}
M.~L. Mehta,
\newblock \emph{{Random Matrices}},
\newblock ISSN. Elsevier Science,
\newblock ISBN 9780080474113 (2004).

\bibitem{Haake2010}
F.~Haake,
\newblock \emph{{Quantum signatures of chaos}},
\newblock Springer series in synergetics. Springer, Berlin [u.a.],
\newblock ISBN 9783642054273 (2010).

\bibitem{French1970}
J.~B. French and S.~S.~M. Wong,
\newblock \emph{{Validity of random matrix theories for many-particle systems}},
\newblock Phys. Lett. B \textbf{33}, 449 (1970),
\newblock \doi{10.1016/0370-2693(70)90213-3}.

\bibitem{BOHIGAS1971}
O.~Bohigas and J.~Flores,
\newblock \emph{Two-body random hamiltonian and level density},
\newblock Physics Letters B \textbf{34}(4), 261 (1971),
\newblock \doi{https://doi.org/10.1016/0370-2693(71)90598-3}.

\bibitem{Dyson1972}
F.~J. Dyson,
\newblock \emph{{A class of matrix ensembles}},
\newblock J. Math. Phys. \textbf{13}, 90 (1972),
\newblock \doi{10.1063/1.1665857}.

\bibitem{Wong1972}
S.~S.~M. Wong and J.~B. French,
\newblock \emph{{LEVEL DENSITY FLUCTUATIONS AND TWO-BODY vs MULTIBODY INTERACTIONS}},
\newblock Nucl. Phys. A \textbf{198}, 188 (1972),
\newblock \doi{10.1016/0375-9474(72)90779-8}.

\bibitem{Mon1975}
K.~K. Mon and J.~B. French,
\newblock \emph{{Statistical Properties of Many Particle Spectra}},
\newblock Annals Phys. \textbf{95}, 90 (1975),
\newblock \doi{10.1016/0003-4916(75)90045-7}.

\bibitem{Brody1981}
T.~A. Brody, J.~Flores, J.~B. French, P.~A. Mello, A.~Pandey and S.~S.~M. Wong,
\newblock \emph{{Random matrix physics: Spectrum and strength fluctuations}},
\newblock Rev. Mod. Phys. \textbf{53}, 385 (1981),
\newblock \doi{10.1103/RevModPhys.53.385}.

\bibitem{Draayer1977}
J.~P. Draayer, J.~B. French and S.~S.~M. Wong,
\newblock \emph{Strength distributions and statistical spectroscopy. {I}. {G}eneral theory},
\newblock Ann.~Phys. \textbf{106}(2), 472 (1977),
\newblock \doi{https://doi.org/10.1016/0003-4916(77)90321-9}.

\bibitem{French1987}
J.~B. French, V.~K.~B. Kota, A.~Pandey and S.~Tomsovic,
\newblock \emph{{Statistical Properties of Many Particle Spectra {VI}. {F}luctuation Bounds on $N-N$ {T-N}oninvariance}},
\newblock Ann.~Phys. \textbf{181}, 235 (1988),
\newblock \doi{10.1016/0003-4916(88)90166-2}.

\bibitem{Tomsovic1986}
S.~Tomsovic,
\newblock \emph{Bounds on the Time-Reversal Noninvariant Nucleon-Nucleon Interaction Derived from Transition Strength Fluctuations},
\newblock Ph.D. thesis, University of Rochester,
\newblock Report number 974, 1987 (1986).

\bibitem{Kota2014}
V.~K.~B. Kota,
\newblock \emph{{Embedded Random Matrix Ensembles in Quantum Physics}},
\newblock Springer,
\newblock \doi{https://doi.org/10.1007/978-3-319-04567-2} (2014).

\bibitem{Benet2002}
L.~Benet and H.~A. Weidenmuller,
\newblock \emph{{Review of the k body embedded ensembles of Gaussian random matrices}},
\newblock J. Phys. A \textbf{36}, 3569 (2003),
\newblock \doi{10.1088/0305-4470/36/12/340},
\newblock \eprint{cond-mat/0207656}.

\bibitem{Kota2001}
V.~Kota,
\newblock \emph{Embedded random matrix ensembles for complexity and chaos in finite interacting particle systems},
\newblock Physics Reports \textbf{347}(3), 223 (2001),
\newblock \doi{https://doi.org/10.1016/S0370-1573(00)00113-7}.

\bibitem{Kota:2018}
V.~K.~B. Kota and N.~D. Chavda,
\newblock \emph{{Embedded random matrix ensembles from nuclear structure and their recent applications}},
\newblock Int. J. Mod. Phys. E \textbf{27}(01), 1830001 (2018),
\newblock \doi{10.1142/S0218301318300011}.

\bibitem{CHANG1971}
F.~Chang, J.~French and T.~Thio,
\newblock \emph{Distribution methods for nuclear energies, level densities, and excitation strengths},
\newblock Annals of Physics \textbf{66}(1), 137 (1971),
\newblock \doi{https://doi.org/10.1016/0003-4916(71)90186-2}.

\bibitem{Kitaev2016_1}
A.~Kitaev,
\newblock \emph{A simple model of quantum holography (part 1)},
\newblock talk at KITP \url{http://online.kitp.ucsb.edu/online/entangled15/kitaev/} (April 7, 2015).

\bibitem{Kitaev2016_2}
A.~Kitaev,
\newblock \emph{{A simple model of quantum holography (part 2)}},
\newblock talk at KITP \url{http://online.kitp.ucsb.edu/online/entangled15/kitaev2/} (May 27, 2015).

\bibitem{Maldacena2016b}
J.~Maldacena and D.~Stanford,
\newblock \emph{{Remarks on the Sachdev-Ye-Kitaev model}},
\newblock Physical Review D \textbf{94}(10) (2016),
\newblock \doi{10.1103/PhysRevD.94.106002},
\newblock \eprint{1604.07818}.

\bibitem{garcia_2016}
A.~M. Garc\'{\i}a-Garc\'{\i}a and J.~J.~M. Verbaarschot,
\newblock \emph{Spectral and thermodynamic properties of the {S}achdev-{Y}e-{K}itaev model},
\newblock Phys. Rev. D \textbf{94}, 126010 (2016),
\newblock \doi{10.1103/PhysRevD.94.126010}.

\bibitem{Garcia_2017}
A.~M. García-García and J.~J.~M. Verbaarschot,
\newblock \emph{Analytical spectral density of the {S}achdev-{Y}e-{K}itaev model at finite {$N$}},
\newblock Phys.~Rev.~D \textbf{96}(6) (2017),
\newblock \doi{10.1103/physrevd.96.066012}.

\bibitem{Cotler2017}
J.~S. Cotler, G.~Gur-Ari, M.~Hanada, J.~Polchinski, P.~Saad, S.~H. Shenker, D.~Stanford, A.~Streicher and M.~Tezuka,
\newblock \emph{{Black holes and random matrices}},
\newblock Journal of High Energy Physics \textbf{2017}(5) (2017),
\newblock \doi{10.1007/JHEP05(2017)118},
\newblock \eprint{1611.04650}.

\bibitem{Maldacena2015}
J.~Maldacena, S.~H. Shenker and D.~Stanford,
\newblock \emph{{A bound on chaos}},
\newblock Journal of High Energy Physics \textbf{2016}(8) (2016),
\newblock \doi{10.1007/JHEP08(2016)106},
\newblock \eprint{1503.01409}.

\bibitem{Kobrin2020}
B.~Kobrin, Z.~Yang, G.~D. Kahanamoku-Meyer, C.~T. Olund, J.~E. Moore, D.~Stanford and N.~Y. Yao,
\newblock \emph{{Many-Body Chaos in the Sachdev-Ye-Kitaev Model}},
\newblock Physical Review Letters \textbf{126}(3), 030602 (2021),
\newblock \doi{10.1103/PhysRevLett.126.030602},
\newblock \eprint{2002.05725}.

\bibitem{Berkooz:2018jqr}
M.~Berkooz, M.~Isachenkov, V.~Narovlansky and G.~Torrents,
\newblock \emph{{Towards a full solution of the large N double-scaled SYK model}},
\newblock JHEP \textbf{03}, 079 (2019),
\newblock \doi{10.1007/JHEP03(2019)079},
\newblock \eprint{1811.02584}.

\bibitem{Baur_2023}
B.~Mukhametzhanov,
\newblock \emph{{Large p SYK from chord diagrams}},
\newblock JHEP \textbf{09}, 154 (2023),
\newblock \doi{10.1007/JHEP09(2023)154},
\newblock \eprint{2303.03474}.

\bibitem{Maldacena2016}
J.~Maldacena, D.~Stanford and Z.~Yang,
\newblock \emph{{Conformal symmetry and its breaking in two-dimensional nearly anti-de Sitter space}},
\newblock Progress of Theoretical and Experimental Physics \textbf{2016}(12) (2016),
\newblock \doi{10.1093/ptep/ptw124},
\newblock \eprint{1606.01857}.

\bibitem{Jensen2016}
K.~Jensen,
\newblock \emph{{Chaos in $AdS_2$ Holography}},
\newblock Physical Review Letters \textbf{117}(11), 111601 (2016),
\newblock \doi{10.1103/PhysRevLett.117.111601},
\newblock \eprint{1605.06098}.

\bibitem{Mertens2023}
T.~G. Mertens and G.~J. Turiaci,
\newblock \emph{Solvable models of quantum black holes: a review on jackiw--teitelboim gravity},
\newblock Living Reviews in Relativity \textbf{26}(1), 4 (2023),
\newblock \doi{10.1007/s41114-023-00046-1}.

\bibitem{Erdos2014}
L.~Erdős and D.~Schröder,
\newblock \emph{Phase transition in the density of states of quantum spin glasses},
\newblock Mathematical Physics, Analysis and Geometry \textbf{17}(3–4), 441–464 (2014),
\newblock \doi{10.1007/s11040-014-9164-3}.

\bibitem{Berkooz:2018qkz}
M.~Berkooz, P.~Narayan and J.~Simon,
\newblock \emph{{Chord diagrams, exact correlators in spin glasses and black hole bulk reconstruction}},
\newblock JHEP \textbf{08}, 192 (2018),
\newblock \doi{10.1007/JHEP08(2018)192},
\newblock \eprint{1806.04380}.

\bibitem{Vyas2018}
M.~Vyas and V.~K.~B. Kota,
\newblock \emph{{Quenched many-body quantum dynamics with k-body interactions using q-Hermite polynomials}},
\newblock J. Stat. Mech. \textbf{1910}(10), 103103 (2019),
\newblock \doi{10.1088/1742-5468/ab4180},
\newblock \eprint{1805.00636}.

\bibitem{Berkooz:2024}
M.~Berkooz and O.~Mamroud,
\newblock \emph{{A cordial introduction to double scaled SYK}},
\newblock Rept. Prog. Phys. \textbf{88}(3), 036001 (2025),
\newblock \doi{10.1088/1361-6633/ada889},
\newblock \eprint{2407.09396}.

\bibitem{Baldwin:2019}
C.~L. Baldwin and B.~Swingle,
\newblock \emph{{Quenched vs Annealed: Glassiness from SK to SYK}},
\newblock Phys. Rev. X \textbf{10}(3), 031026 (2020),
\newblock \doi{10.1103/PhysRevX.10.031026},
\newblock \eprint{1911.11865}.

\bibitem{Swingle_2024}
B.~Swingle and M.~Winer,
\newblock \emph{Bosonic model of quantum holography},
\newblock Phys. Rev. B \textbf{109}, 094206 (2024),
\newblock \doi{10.1103/PhysRevB.109.094206}.

\bibitem{Jia:2024tii}
Y.~Jia,
\newblock \emph{{Bosonic near-CFT$_{1}$ models from Fock-space fluxes}},
\newblock JHEP \textbf{07}, 162 (2024),
\newblock \doi{10.1007/JHEP07(2024)162},
\newblock \eprint{2403.13131}.

\bibitem{Lin:2022rbf}
H.~W. Lin,
\newblock \emph{{The bulk Hilbert space of double scaled SYK}},
\newblock JHEP \textbf{11}, 060 (2022),
\newblock \doi{10.1007/JHEP11(2022)060},
\newblock \eprint{2208.07032}.

\bibitem{Sachdev:2015efa}
S.~Sachdev,
\newblock \emph{{Bekenstein-Hawking Entropy and Strange Metals}},
\newblock Phys. Rev. X \textbf{5}(4), 041025 (2015),
\newblock \doi{10.1103/PhysRevX.5.041025},
\newblock \eprint{1506.05111}.

\bibitem{sachdev_1993}
S.~Sachdev and J.~Ye,
\newblock \emph{Gapless spin-fluid ground state in a random quantum heisenberg magnet},
\newblock Phys. Rev. Lett. \textbf{70}, 3339 (1993),
\newblock \doi{10.1103/PhysRevLett.70.3339}.

\bibitem{Georges2001}
A.~Georges, O.~Parcollet and S.~Sachdev,
\newblock \emph{Quantum fluctuations of a nearly critical heisenberg spin glass},
\newblock Phys. Rev. B \textbf{63}, 134406 (2001),
\newblock \doi{10.1103/PhysRevB.63.134406}.

\bibitem{Parcollet_1998}
O.~Parcollet, A.~Georges, G.~Kotliar and A.~Sengupta,
\newblock \emph{{Overscreened multi-channel $SU(N)$ Kondo model: Large-$N$ solution and conformal field theory}},
\newblock Physical Review B \textbf{58}(7), 3794–3813 (1998),
\newblock \doi{10.1103/physrevb.58.3794}.

\bibitem{Sachdev_2010a}
S.~Sachdev,
\newblock \emph{Holographic metals and the fractionalized fermi liquid},
\newblock Physical Review Letters \textbf{105}(15) (2010),
\newblock \doi{10.1103/physrevlett.105.151602}.

\bibitem{Sachdev_2010b}
S.~Sachdev,
\newblock \emph{Strange metals and the ads/cft correspondence},
\newblock Journal of Statistical Mechanics: Theory and Experiment \textbf{2010}(11), P11022 (2010),
\newblock \doi{10.1088/1742-5468/2010/11/P11022}.

\bibitem{Vyas2012}
M.~Vyas and V.~K.~B. Kota,
\newblock \emph{{Embedded Gaussian Unitary Ensembles with $U(Ω) \otimes SU(r)$ Embedding generated by Random Two-body Interactions with $SU(r)$ Symmetry}}  (2012),
\newblock \doi{10.1063/1.4768711},
\newblock \eprint{1207.7032}.

\bibitem{Gu2020}
Y.~Gu, A.~Kitaev, S.~Sachdev and G.~Tarnopolsky,
\newblock \emph{{Notes on the complex Sachdev-Ye-Kitaev model}},
\newblock Journal of High Energy Physics \textbf{2020}(2) (2020),
\newblock \doi{10.1007/jhep02(2020)157},
\newblock \eprint{1910.14099}.

\bibitem{Davison_2017}
R.~A. Davison, W.~Fu, A.~Georges, Y.~Gu, K.~Jensen and S.~Sachdev,
\newblock \emph{Thermoelectric transport in disordered metals without quasiparticles: The sachdev-ye-kitaev models and holography},
\newblock Physical Review B \textbf{95}(15) (2017),
\newblock \doi{10.1103/physrevb.95.155131}.

\bibitem{sachdev2021}
M.~Tikhanovskaya, H.~Guo, S.~Sachdev and G.~Tarnopolsky,
\newblock \emph{Excitation spectra of quantum matter without quasiparticles. i. sachdev-ye-kitaev models},
\newblock Phys. Rev. B \textbf{103}, 075141 (2021),
\newblock \doi{10.1103/PhysRevB.103.075141}.

\bibitem{Berkooz:2020}
M.~Berkooz, V.~Narovlansky and H.~Raj,
\newblock \emph{{Complex Sachdev-Ye-Kitaev model in the double scaling limit}},
\newblock JHEP \textbf{02}, 113 (2021),
\newblock \doi{10.1007/JHEP02(2021)113},
\newblock \eprint{2006.13983}.

\bibitem{riordan_1975}
J.~Riordan,
\newblock \emph{The distribution of crossings of chords joining pairs of $2n$ points on a circle},
\newblock Mathematics of Computation \textbf{29}(129), 215 (1975).

\bibitem{Touchard_1952}
J.~Touchard,
\newblock \emph{Sur un problème de configurations et sur les fractions continues},
\newblock Canadian Journal of Mathematics \textbf{4}, 2–25 (1952),
\newblock \doi{10.4153/CJM-1952-001-8}.

\bibitem{avram1987}
F.~Avram and M.~S. Taqqu,
\newblock \emph{{Noncentral Limit Theorems and Appell Polynomials}},
\newblock The Annals of Probability \textbf{15}(2), 767  (1987),
\newblock \doi{10.1214/aop/1176992170}.

\bibitem{ISMAIL1987}
M.~E. Ismail, D.~Stanton and G.~Viennot,
\newblock \emph{The combinatorics of q-hermite polynomials and the askey—wilson integral},
\newblock European Journal of Combinatorics \textbf{8}(4), 379 (1987),
\newblock \doi{https://doi.org/10.1016/S0195-6698(87)80046-X}.

\bibitem{Vyas_2020}
M.~Vyas and V.~K.~B. Kota,
\newblock \emph{Bivariate q-normal distribution for transition matrix elements in quantum many-body systems},
\newblock Journal of Statistical Mechanics: Theory and Experiment \textbf{2020}(9), 093101 (2020),
\newblock \doi{10.1088/1742-5468/ababfc}.

\bibitem{Kota2022}
V.~K.~B. Kota and M.~Vyas,
\newblock \emph{{Statistical nuclear spectroscopy with q-normal and bivariate q-normal distributions and q-hermite polynomials}},
\newblock Annals Phys. \textbf{446}, 169131 (2022),
\newblock \doi{10.1016/j.aop.2022.169131},
\newblock \eprint{2201.12509}.

\bibitem{szabowski2013}
P.~J. Szabłowski,
\newblock \emph{On the q-hermite polynomials and their relationship with some other families of orthogonal polynomials},
\newblock Demonstratio Mathematica \textbf{46}, 679  (2011),
\newblock \doi{https://doi.org/10.48550/arXiv.1101.2875}.

\bibitem{D_Alessio_2016}
L.~D’Alessio, Y.~Kafri, A.~Polkovnikov and M.~Rigol,
\newblock \emph{From quantum chaos and eigenstate thermalization to statistical mechanics and thermodynamics},
\newblock Advances in Physics \textbf{65}(3), 239–362 (2016),
\newblock \doi{10.1080/00018732.2016.1198134}.

\bibitem{wurm2002}
A.~Wurm and M.~Berg,
\newblock \emph{Wick calculus}  (2002),
\newblock \doi{https://doi.org/10.48550/arXiv.physics/0212061},
\newblock \eprint{physics/0212061}.

\bibitem{Bhattacharya_2017}
R.~Bhattacharya, S.~Chakrabarti, D.~P. Jatkar and A.~Kundu,
\newblock \emph{Syk model, chaos and conserved charge},
\newblock Journal of High Energy Physics \textbf{2017}(11) (2017),
\newblock \doi{10.1007/jhep11(2017)180}.

\bibitem{Lin2023}
H.~W. Lin and D.~Stanford,
\newblock \emph{{A symmetry algebra in double-scaled SYK}},
\newblock SciPost Phys. \textbf{15}, 234 (2023),
\newblock \doi{10.21468/SciPostPhys.15.6.234}.

\bibitem{Eynard2018}
B.~Eynard, T.~Kimura and S.~Ribault,
\newblock \emph{{Random matrices}}  (2018),
\newblock \doi{https://doi.org/10.48550/arXiv.1510.04430},
\newblock \eprint{1510.04430}.

\bibitem{Saad2019}
P.~Saad, S.~H. Shenker and D.~Stanford,
\newblock \emph{{JT gravity as a matrix integral}}  (2019),
\newblock \doi{https://doi.org/10.48550/arXiv.1903.11115},
\newblock \eprint{1903.11115}.

\bibitem{Berkooz:2020fvm}
M.~Berkooz, N.~Brukner, V.~Narovlansky and A.~Raz,
\newblock \emph{{Multi-trace correlators in the SYK model and non-geometric wormholes}},
\newblock JHEP \textbf{21}, 196 (2020),
\newblock \doi{10.1007/JHEP09(2021)196},
\newblock \eprint{2104.03336}.

\bibitem{Altland:2017eao}
A.~Altland and D.~Bagrets,
\newblock \emph{{Quantum ergodicity in the SYK model}},
\newblock Nucl. Phys. B \textbf{930}, 45 (2018),
\newblock \doi{10.1016/j.nuclphysb.2018.02.015},
\newblock \eprint{1712.05073}.

\bibitem{Kota2023}
V.~K.~B. Kota,
\newblock \emph{Bivariate moments of the two-point correlation function for embedded gaussian unitary ensemble with $k$-body interactions},
\newblock Phys. Rev. E \textbf{107}, 054128 (2023),
\newblock \doi{10.1103/PhysRevE.107.054128}.

\bibitem{Goel:2023svz}
A.~Goel, V.~Narovlansky and H.~Verlinde,
\newblock \emph{{Semiclassical geometry in double-scaled SYK}},
\newblock JHEP \textbf{11}, 093 (2023),
\newblock \doi{10.1007/JHEP11(2023)093},
\newblock \eprint{2301.05732}.

\bibitem{askey1994}
R.~A. Askey, M.~Rahman and S.~Suslov,
\newblock \emph{On a general q-fourier transformation with nonsymmetric kernels},
\newblock Journal of Computational and Applied Mathematics \textbf{68}(1), 25 (1996),
\newblock \doi{https://doi.org/10.1016/0377-0427(95)00259-6}.

\bibitem{Szabowski2011}
P.~J. Szabłowski,
\newblock \emph{On the structure and probabilistic interpretation of askey–wilson densities and polynomials with complex parameters},
\newblock Journal of Functional Analysis \textbf{261}(3), 635–659 (2011),
\newblock \doi{10.1016/j.jfa.2011.04.002}.

\end{thebibliography}


\end{document}